\documentclass[aip, amsmath,amssymb, preprint,]{revtex4-1}

\usepackage{amsmath}
\usepackage{graphicx}
\usepackage{multirow}
\usepackage{array}
\usepackage{color} 
\usepackage{booktabs}
\usepackage{amsfonts}
\usepackage{hyperref}
\usepackage{textcomp}
\usepackage{ulem}
\usepackage[T1]{fontenc}
\usepackage{mathptmx}
\usepackage[utf8]{inputenc}
\usepackage{lineno}

\begin{document}

\title {Orbitals of Artificial Atoms in a Gapped Two-Dimensional Vacuum}

\author {Mong-Wen Gu}
\affiliation {Peter Gr\"{u}nberg Institut (PGI-3), Forschungszentrum J\"{u}lich, 52425 J\"{u}lich, Germany}
\affiliation{J\"ulich Aachen Research Alliance (JARA), Fundamentals of Future Information Technology, 52425 J\"ulich, Germany}
\author {Aizhan Sabitova}
\affiliation {Peter Gr\"{u}nberg Institut (PGI-3), Forschungszentrum J\"{u}lich, 52425 J\"{u}lich, Germany}
\affiliation{J\"ulich Aachen Research Alliance (JARA), Fundamentals of Future Information Technology, 52425 J\"ulich, Germany}
\author {Taner Esat}
\affiliation {Peter Gr\"{u}nberg Institut (PGI-3), Forschungszentrum J\"{u}lich, 52425 J\"{u}lich, Germany}
\affiliation{J\"ulich Aachen Research Alliance (JARA), Fundamentals of Future Information Technology, 52425 J\"ulich, Germany}
\author {Christian Wagner}
\affiliation {Peter Gr\"{u}nberg Institut (PGI-3), Forschungszentrum J\"{u}lich, 52425 J\"{u}lich, Germany}
\affiliation{J\"ulich Aachen Research Alliance (JARA), Fundamentals of Future Information Technology, 52425 J\"ulich, Germany}
\author{F. Stefan Tautz}
\affiliation {Peter Gr\"{u}nberg Institut (PGI-3), Forschungszentrum J\"{u}lich, 52425 J\"{u}lich, Germany}
\affiliation{J\"ulich Aachen Research Alliance (JARA), Fundamentals of Future Information Technology, 52425 J\"ulich, Germany}
\affiliation{Experimentalphysik IV A, RWTH Aachen University, 52074 Aachen, Germany}
\author {Aleksandr Rodin}
\affiliation {Department of Materials Science and Engineering, National University of Singapore, Singapore 117575, Singapore}
\author {Ruslan Temirov}
\email[corresponding author: ]{r.temirov@fz-juelich.de}
\affiliation {Peter Gr\"{u}nberg Institut (PGI-3), Forschungszentrum J\"{u}lich, 52425 J\"{u}lich, Germany}
\affiliation{J\"ulich Aachen Research Alliance (JARA), Fundamentals of Future Information Technology, 52425 J\"ulich, Germany}
\affiliation{University of Cologne, Faculty of Mathematics and Natural Sciences, Institute of Physics II, 50937 Cologne, Germany}

\begin{abstract}

Advances in nanotechnology now allow the creation of artificial atoms -- engineered structures whose electronic states closely mimic those of real atoms. 
Understanding how these artificial atoms interact and bond is key to designing new materials with tailored electronic properties. 
Here, we use scanning tunnelling microscopy to visualise the bound states of nanostructures patterned in a two-dimensional molecular film featuring a parabolic band with multiple partial energy gaps. 
The lowest-energy states split off from the bottom of the band and resemble the familiar $s$ and $p$ orbitals of natural atoms, even bonding in the same way. 
Yet, artificial atoms go beyond this analogy: the gapped two-dimensional vacuum in which they reside gives rise to entirely new orbitals with no counterparts in real atoms.
These quasi-one-dimensional localised states enrich the orbital vocabulary of chemistry, adding a new class of orbitals that are predominantly shaped by the surrounding electronic vacuum.

\end{abstract}

\maketitle

\section{Introduction}

Modern chemistry owes much of its success to the picture of atomic orbitals.
Although this picture is a mere approximation of the multi-electron states in atoms and molecules \cite{pham2017JPC,kumar2025JACS}, it proves its value in making chemistry intuitive \cite{fukui1977orbital,stowasser_what_1999}, visualising it as a coupling of atomic orbitals induced by their spatial overlap. 
This intuition may again become helpful as progress in nanotechnology and nanofabrication leads to the development of "artificial atoms" -- systems whose electronic states resemble those of natural atoms \cite{kastner_artificial_1993, tarucha_electronic_1998, crommie_confinement_1993, Abajo2013, folsch_quantum_2014, peng_visualizing_2021, Freeney_coupling_coralls_2020, pan_reconfigurable_2015, jolie_creating_2022, Sierda2023Science}.
In particular, it could aid in designing new artificial materials featuring interesting electronic properties \cite{Gomes2012, Drost2017, Slot2017, slot_p_2019, kempkes_design_2019, kempkes_robust_2019, khajetoorians_creating_2019, pham_topological_2022, freeney_electronic_2022,fang_atomically-precise_2023, Sierda2023Science}.
To assess how far the atomic orbital picture extends to artificial atoms, it is helpful to break it up into its conceptual ingredients.
Orbitals arise as bound-state solutions of an attractive atomic potential in a 3D vacuum in which electrons move freely.
Thus, the key factors are the dimensionality, the shape of the potential, and the dispersion of a free electron in a vacuum.
Not surprisingly, the artificial atoms may differ from their natural counterparts in all of the above characteristics.
Artificial atom setups are typically prepared on a surface and are thus two-dimensional \cite{crommie_confinement_1993, Gomes2012, Abajo2013, folsch_quantum_2014, pan_reconfigurable_2015, Drost2017, Slot2017, slot_p_2019, kempkes_design_2019, kempkes_robust_2019, Freeney_coupling_coralls_2020, peng_visualizing_2021,  khajetoorians_creating_2019, jolie_creating_2022, pham_topological_2022, freeney_electronic_2022,fang_atomically-precise_2023, Sierda2023Science}.
The attractive potential is usually shallow and flat-bottomed, and it also lacks ideal symmetry \cite{folsch_quantum_2014, pan_reconfigurable_2015, Sierda2023Science}.
Finally, they always exist in a material rather than a pristine vacuum, and therefore the host's electronic structure defines the dispersion, serving as a new "vacuum".

Here, we utilise a low-temperature scanning tunnelling microscope (STM) to create artificial nanostructures that exert an attractive potential on electrons propagating in a two-dimensional (2D) electron band, which realises the essential ingredients of the 2D artificial atom concept. 
The special advantage of our approach lies in combining a novel imaging method capable of resolving individual orbitals with precise knowledge of the 2D band dispersion that defines the system's electronic "vacuum".
Close to the $\bar\Gamma$ point, the 2D band exhibits an almost parabolic dispersion, and we discover by direct imaging that our artificial atoms possess analogues of $s$ and $p$ orbitals that energetically split from the parabolic bottom of the 2D band.
By creating pairs of atoms, we further illustrate how their orbitals couple when they overlap.
Because our artificial atoms reside in a 2D vacuum with multiple energy gaps induced by a periodically corrugated potential, they exhibit a new type of orbital not present in natural atoms. 
These are quasi-one-dimensional states lying inside the 2D band energy gaps above the "vacuum level" defined by the onset of the 2D band at the $\bar\Gamma$ point.

\section{Results}

\subsection{Artificial atom $s$ and $p$ orbitals}

\begin{figure}
\centering
\includegraphics[width=14cm]{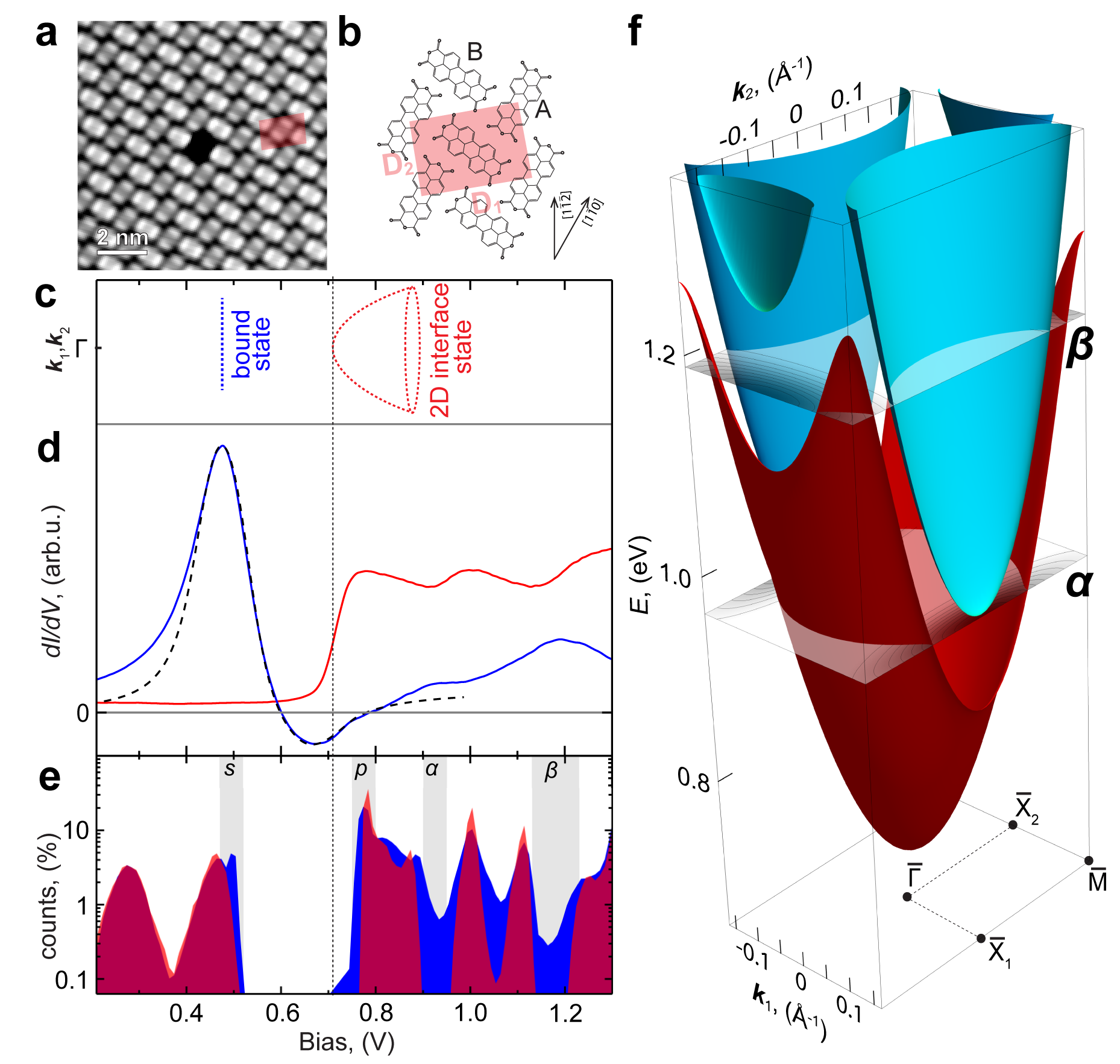}

\caption{\textbf{Bound-state formation at a molecular vacancy.} 
(\textbf{a}) STM image of a type-B vacancy in a commensurate PTCDA/Ag(111) monolayer (here and in all other images, scale bar: 2 nm). The dashed rectangle marks the molecular unit cell with inequivalent A and B molecules. 
(\textbf{b}) Schematic of the monolayer structure in the same orientation; arrows indicate high-symmetry directions of Ag(111). 
(\textbf{c}) Illustration of a 2D band with a bound state split from its parabolic bottom. 
(\textbf{d}) $dI/dV$ spectra recorded above a clean surface (red) and a vacancy (blue), with a Newns-Anderson fit (black dashed) identifying the bound state. 
(\textbf{e}) Energy-distribution histograms (EDHs; see Methods) from a defect-free area (red) and the area in (a) (blue), highlighting vacancy-induced states (grey bands) used for feature-distribution maps (FDMs; see Methods). 
(\textbf{f}) Calculated band structure of the PTCDA/Ag(111) 2D interface-state (see also Fig. S3). Horizontal planes mark the energies of the $\alpha$ and $\beta$ states observed in the experiments. Each plane additionally exhibits the Gaussian-shaped $\mathbf{k}$ vector distributions used to obtain the densities in Fig.~\ref{vacB_FDSTS_u}e-f}.

\label{vacB_STS} 
\end{figure}

We engineer our artificial atom structures by creating individual vacancies (Fig.~\ref{vacB_STS}a) in a commensurate monolayer of Perylene-3,4,9,10-tetracarboxylic dianhydride (PTCDA) molecules chemisorbed on an Ag(111) surface \cite{Glockler1998SS, Rohlfing2007, tautz_structure_2007}. 
The unit cell of the herringbone PTCDA/Ag(111) monolayer features two inequivalent molecules, A (bright) and B (dark) (Fig.~\ref{vacB_STS}b), allowing for two distinct types of vacancies. 
We prepare a type-B vacancy by removing a B molecule using the STM tip. \cite{temirov_kondo_2008, Toher2011, green_patterning_2014, leinen_virtual_2015}
Note that although we exclusively discuss type-B vacancies here, the behaviour of type-A vacancies is similar.

Because PTCDA/Ag(111) possesses a hybrid 2D interface state \cite{Temirov2006Nature, Schwalb2008PRL, Dyer2010NJP, Galbraith2014, Sabitova2018PRB}, the artificial electronic states discussed here are due to the action of the vacancy potential onto the 2D interface state, serving here as an electronic "vacuum" (Fig.\ref{vacB_STS}c). 
The onset of the 2D interface state lies at 710 meV above the Fermi level of Ag(111), rendering it unoccupied (Figs.~\ref{vacB_STS}d,f). 
Thus, all vacancy states are unoccupied and can be observed in STM only by tunnelling individual electrons from the tip into them. 
Because of the relatively strong coupling to the substrate, the probed states never contain more than one electron, thus closely approaching the concept of single-electron states (the lifetime of the tunnelling electron in this state is shorter than the time between successive tunnelling events).
Unlike in real atoms, we therefore do not need to consider electron-electron correlations and many-electron effects. 

Scanning tunnelling spectroscopy (STS) of the type-B vacancy reveals a pronounced resonance, as shown in Fig.~\ref{vacB_STS}d. 
We establish an analogy between the PTCDA vacancy and an atom by demonstrating that the observed STS resonance arises as a bound state in the vacuum surrounding the attractive vacancy potential. 
To this end, we first utilise the modified Newns-Anderson chemisorption model \cite{Limot2005PRL, Kroger2005PSS} (see Methods), which describes the electronic structure of a localised level coupled simultaneously to the 2D surface band and the 3D bulk of a metal. 
A general result of the model is its prediction, under certain conditions, of the appearance of a bound state splitting off the bottom of the 2D band \cite{hammer_special_2006} (Fig.~\ref{vacB_STS}c). 
The fact that the model provides a good fit to our data (Fig.~\ref{vacB_STS}d) indicates that the vacancy resonance may indeed be a bound state of an electron from the 2D interface state band. 
Interestingly, the fit also shows that the bound state has a predominantly 2D character because its coupling to the 2D interface state, $\Delta_s=485$~meV, is substantially higher than to the Ag bulk, $\Delta_b=77$~meV.

As the observed resonance in Fig.~\ref{vacB_STS}d appears to result from the interaction between the vacancy and the 2D interface state, we next map it in space to demonstrate its localisation. 
We note that the electronic states of molecular vacancies have been discussed previously, but unlike the states of quantum corrals \cite{crommie_confinement_1993}, they have never been visualised \cite{Abajo2013}. 
To obtain clear images of the bound states of the molecular vacancy, we apply feature-detection STS (FD STS) -- a powerful approach to the analysis and visualisation of the STS data, which we introduced recently \cite{Sabitova2018PRB, martinez-castro_disentangling_2022}.

\begin{figure}
\centering
\includegraphics[width=8cm]{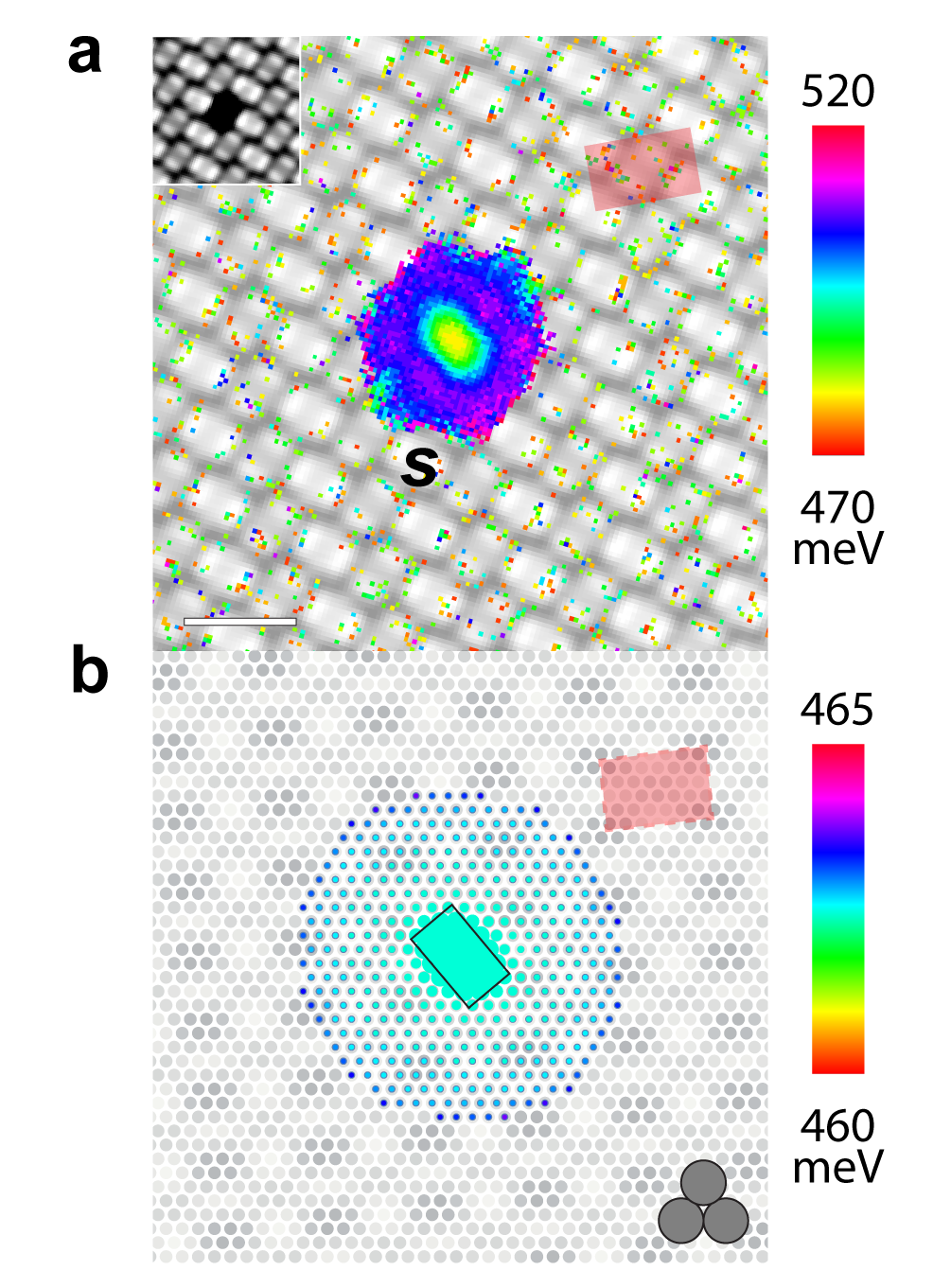}
\caption{\textbf{Artificial $s$ orbital of a single vacancy.} 
(\textbf{a}) STM image of a type-B vacancy overlaid with a feature-detection map (FDM; see Methods) revealing the lowest-energy bound state of the vacancy; made from the portion of the peak-detection statistics marked with "s" in Fig.~\ref{vacB_STS}e.
The colour reflects the peak energy, mapped according to the colorbar next to the image.
The inset provides a zoomed-out view of the vacancy, which in the main panel is covered by the FDM overlay.
(\textbf{b}) FDM obtained from the tight-binding (TB) simulation (see Methods).
The grey corrugation is due to the periodic potential (dark is more attractive).
The black rectangle was used to impose the vacancy potential.
The inset depicts the unit cell of the simulation lattice.
The colour reflects the peak energy, mapped according to the colorbar next to the image.
The size of the coloured nodes reflects the spectral function peak intensity.}

\label{vacB_FDSTS_s} 
\end{figure}

FD STS identifies and maps spectral peaks across a grid of $dI/dV$ spectra, producing energy distribution histograms (EDHs) and feature distribution maps (FDMs). 
For more information, see Methods.  
The two logarithmic EDHs shown in Fig.~\ref{vacB_STS}e stem from two $dI/dV$ grids, one (red) taken over a defect-free PTCDA/Ag(111) area, and the other (blue) over the area with a single type-B vacancy shown in Fig.~\ref{vacB_STS}a. 
The features shared by both EDHs are characteristic of pristine PTCDA/Ag(111). 
The additional features observed in the blue EDH indicate that the type-B vacancy gives rise to new electronic states. 
FDM shown in Fig.~\ref{vacB_FDSTS_s}a provides a spatial map of the lowest energy vacancy state whose EDH is peaking at $\sim$502 meV (marked with "s" in Fig.~\ref{vacB_STS}e). 
The state is localised around but not confined within the vacancy.
It thus behaves as a bound state of the vacancy rather than the states of atomic corrals studied earlier \cite{crommie_confinement_1993, Freeney_coupling_coralls_2020}.

The circular shape of the bound state indicates that the angular momentum quantum number is zero; the orbital, therefore, is a 2D analogue of an $s$ orbital. To substantiate this analogy, we simulate the electronic states of the vacancy using tight-binding (TB) calculations (see  Methods and Supplement). 
To make the output of the calculations comparable to the experiment, we apply FD STS to the simulated spectral functions (see Methods and Supplement) to obtain tight-binding-derived feature-distribution maps (TB FDMs), which show both the energy (node colour) and the intensity (node size) of peaks in the calculated spectral functions across the TB lattice sites. 
A TB FDM shown in Fig.~\ref{vacB_FDSTS_s}b  reproduces very well the experimentally observed bound state, confirming its interpretation as a vacancy-induced orbital.
The energy dispersion (i.e., colour change) in the centre of the experimental FDM and at the very edges of the TB FDM are artefacts of the measurement and the peak-detection algorithm, respectively. 
For a more detailed discussion of these and a few other minor differences between the experimental FDMs and the TB FDMs, see the Methods section.

The good agreement between experiment and simulation confirms that the vacancy indeed acts as an artificial atom.
We next show that it also possesses a $p$ orbital. 
Although this bound state (marked with "p" in Fig.~\ref{vacB_STS}e) lies slightly above the onset of the 2D interface state, we recover its image in the experimental FDM shown in Fig.~\ref{vacB_FDSTS_p}a. 
Because of its shape, we call this state a $p$ orbital. 
The FDM reveals that the $p$ orbital overlaps with the 2D vacuum state energetically but not spatially, because the scattering resonance of the vacuum state is pushed away from the vacancy. 
Note that the strong radial energy dispersion confirms the delocalised nature of the latter (see Methods).
The TB calculation corroborates the experiment: Fig.~\ref{vacB_FDSTS_p}b shows a $p$-like state adjacent to the dispersing feature.

\begin{figure}
\centering
\includegraphics[width=8cm]{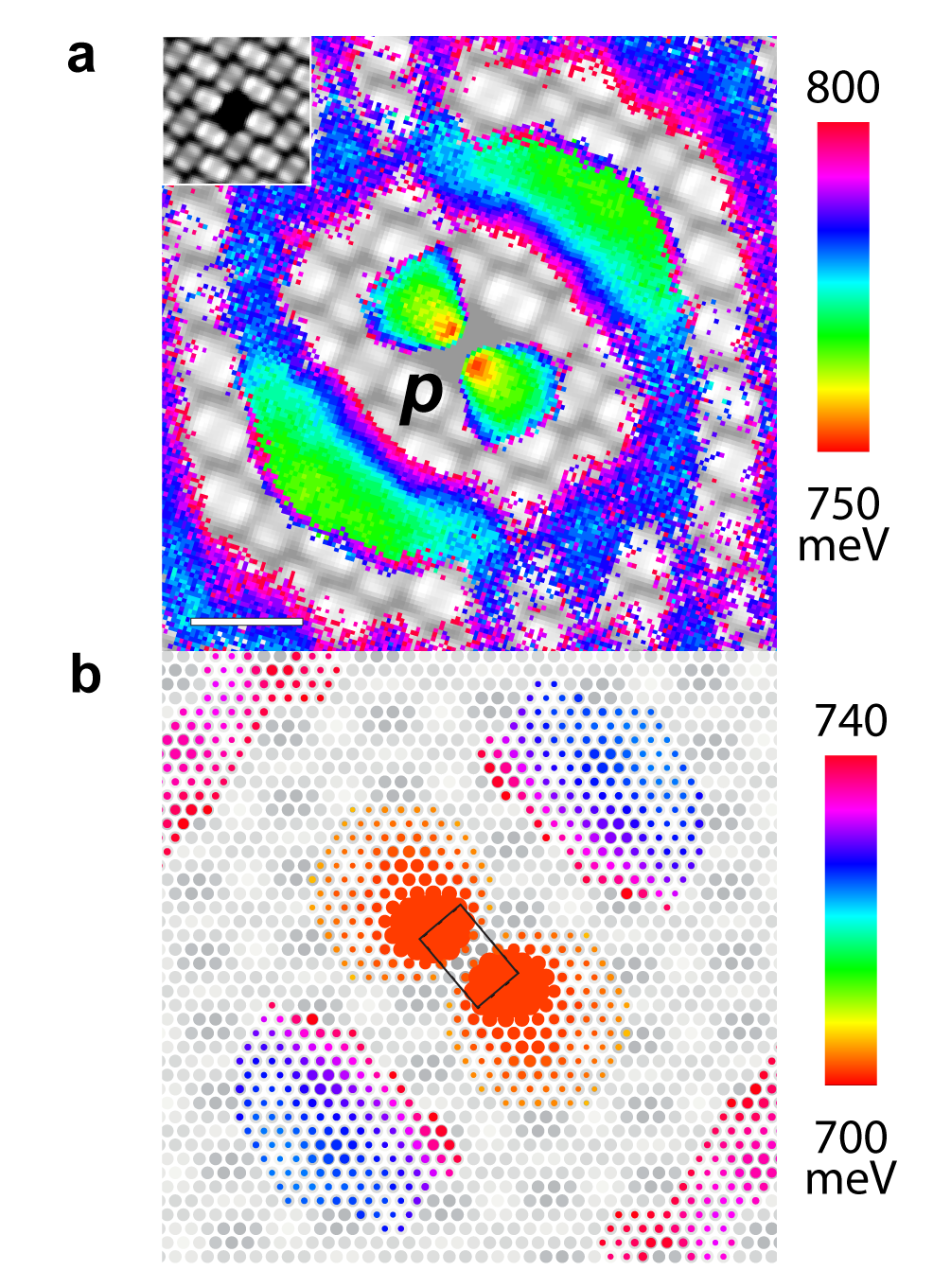}
\caption{\textbf{Artificial $p$ orbital of a single vacancy.} 
(\textbf{a}) STM image of a type-B vacancy overlaid with a feature-detection map (FDM; see Methods) revealing the second-lowest-energy bound state of the vacancy; made from the portion of the peak-detection statistics marked with "p" in Fig.~\ref{vacB_STS}e.
The colour reflects the peak energy, mapped according to the colorbar next to the image.
The inset provides a zoomed-out view of the vacancy, which in the main panel is covered by the FDM overlay.
(\textbf{b}) FDM obtained from the tight-binding (TB) simulation (see Methods).
The grey corrugation is due to the periodic potential (dark - attractive).
The black rectangle was used to impose the vacancy potential.
The inset depicts the unit cell of the simulation lattice.
The colour reflects the peak energy, mapped according to the colorbar next to the image.
The size of the coloured nodes reflects the spectral function peak intensity.}
\label{vacB_FDSTS_p} 
\end{figure}

Both the experiment and simulation find only one $p$ orbital. 
The vacancy potential clearly breaks circular symmetry, lifting the degeneracy between the two possible $p$ orbitals. 
The higher-energy orbital fails to form because its energy is too high.
The scattering resonance could be interpreted as a precursor of this orbital, as one may expect it to smoothly "transform" into the second $p$ orbital by deepening or widening of the vacancy potential.

\subsection{Coupling of artificial atom $s$ and $p$ orbitals}

After visualising the bound states of individual vacancies, we next demonstrate that they also couple, extending the chemical analogy to bonding.
When two type-B nearest neighbour vacancies are patterned 12.6 \AA~apart (inset in Figs.~\ref{vacBB1_FDSTS}a), their $s$ orbitals strongly overlap, producing bonding $\sigma_s$ and antibonding $\sigma_s^*$ hybrids at 395 meV and 590 meV peak value (Fig.~\ref{summary_hybrid}), respectively. 
The resulting coupling strength, $\tau = -98$ meV, agrees well with the tight-binding (TB) simulations (Figs.~\ref{vacBB1_FDSTS}b,d), which spot $\sigma_s$ at 393 meV and $\sigma_s^*$ at 555 meV thus yielding $\tau = -81$ meV.
The nearest-neighbour pair also exhibits a $p$-bybrid orbital whose energy lies deeper in the continuum.
Due to this strong energetic overlap, its analysis is complicated by the limited agreement between theory and experiment (Fig. S8).  

\begin{figure}
\centering
\includegraphics[width=8cm]{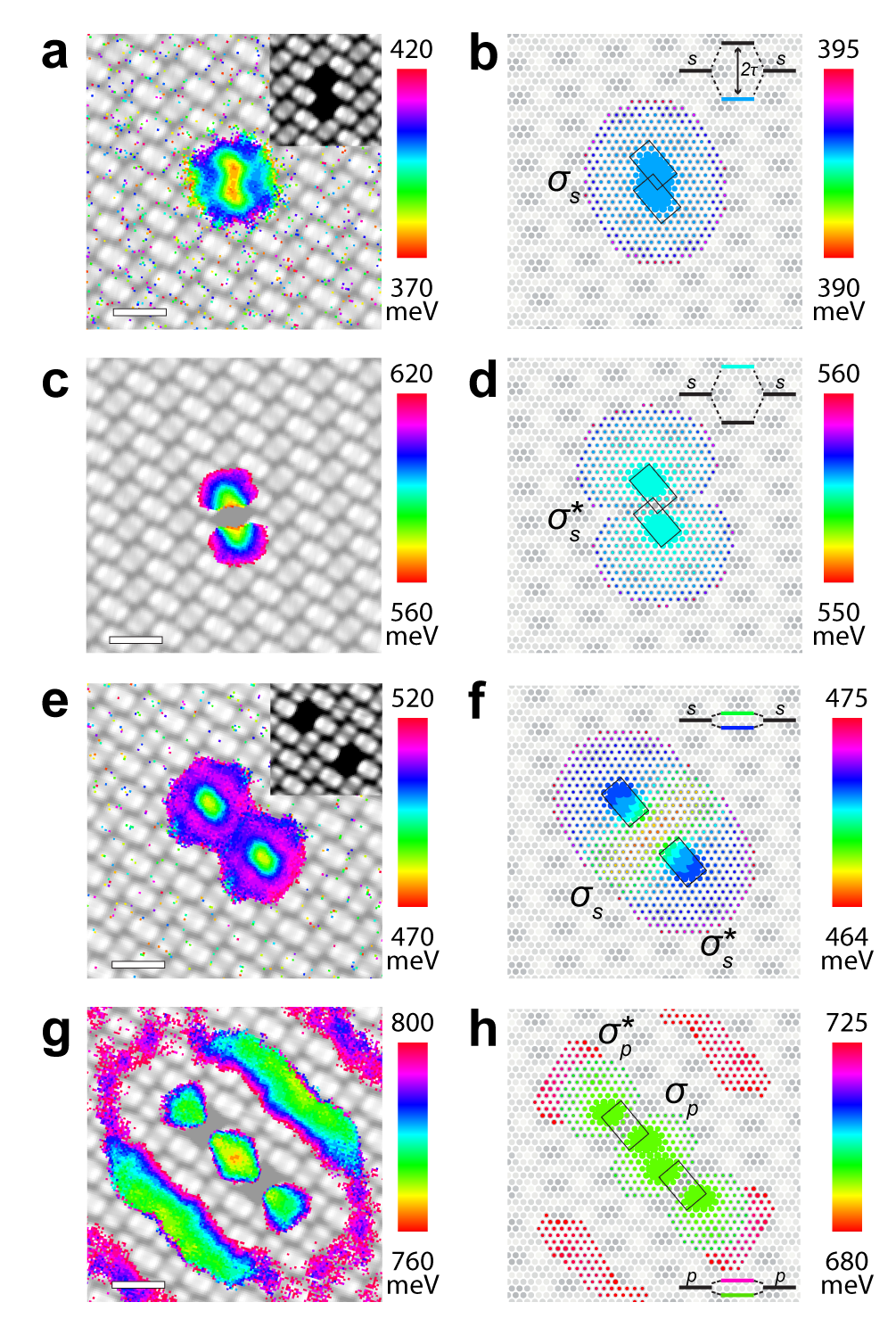}
\caption{\textbf{Coupling of artificial orbitals in vacancy dimers.} 
(\textbf{a--d}) Nearest-neighbour type-B vacancy pair: experimental (a,c) and simulated (b,d) feature-detection maps (FDMs; see Methods) showing the bonding (a,b) and antibonding (c,d) hybrids of two $s$ orbitals. 
(\textbf{e--h}) A weekly interacting type-B vacancy pair: experimental (e,g) and TB simulated (f,h) FDMs highlighting the weak $s-s$ (e-f) and somewhat stronger $p-p$ coupling (g,h).
The black rectangles mark the model vacancy potential (see Methods).
The colour reflects the peak energy, mapped according to the colorbar next to the image.
The size of the coloured nodes reflects the peak intensity.
The insets in (a,e) show the STM topographies of the dimers.
The insets in (b,d,f,h) indicate the degree of coupling and identify the shown hybrid orbital with its FDM colour.
}

\label{vacBB1_FDSTS} 
\end{figure}

Increasing the separation to 31 \AA, as shown in Fig.~\ref{vacBB1_FDSTS}e, suppresses the overlap almost completely: only a weak interaction of $\tau \approx -5$ meV is visible in the TB data (Fig.~\ref{vacBB1_FDSTS}f), consistent with the absence of a measurable energy splitting in the STS due to experimental broadening (see Methods).
The same vacancy pair shows a stronger coupling between their $p$ orbitals (Fig.~\ref{vacBB1_FDSTS}g). 
The experimental feature primarily reveals the bonding hybrid, whereas the TB simulations in Fig.~\ref{vacBB1_FDSTS}h reveal both bonding (green) and antibonding (red) hybrids, separated by about 30 meV and thus suggesting $\tau \approx -15$ meV. 
The antibonding hybrid of the pair energetically overlaps with the scattering resonance of the 2D vacuum. 
Visualising multiorbital coupling in an artificial-atom system in a 2D electronic vacuum, we demonstrate that this coupling depends sensitively on the distance, and, as in real chemistry, on the orbital symmetry. 

\begin{figure}
\centering
\includegraphics[width=8cm]{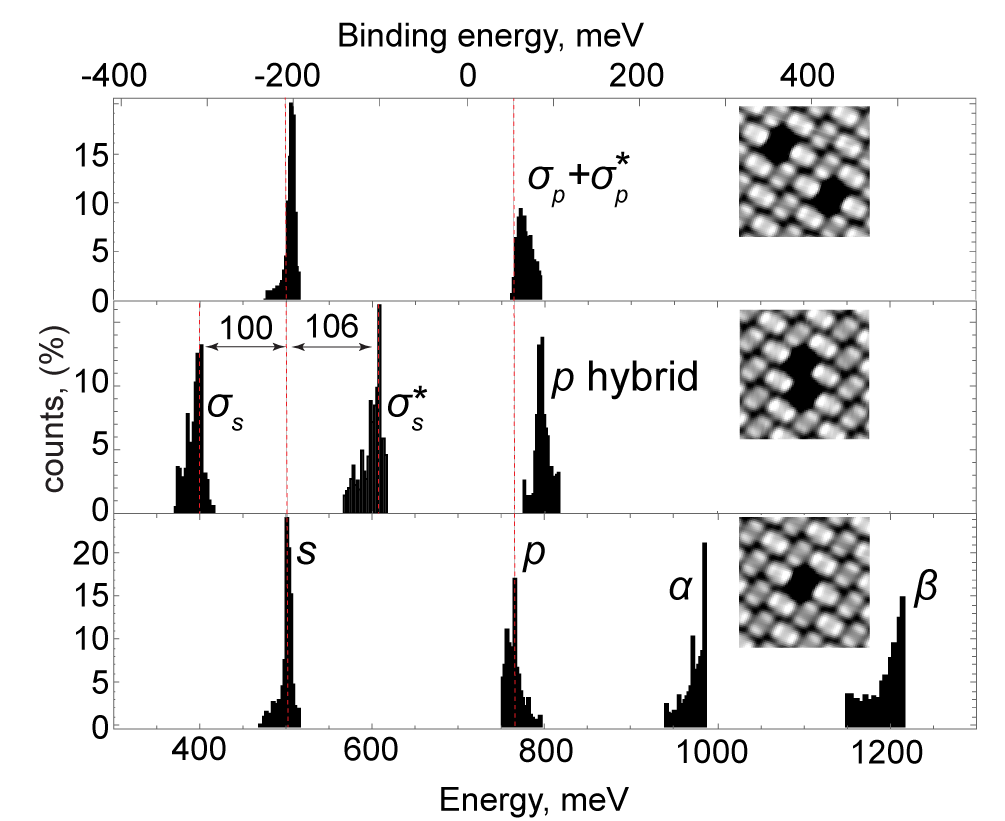}
\caption{\textbf{Evolution of the orbital energy spectrum from isolated to coupled artificial atoms.}
Local energy-distribution histograms (EDHs; see Methods) of individual orbitals of a single type-B vacancy (bottom), a nearest-neighbour dimer (middle), and a weakly interacting vacancy dimer (top).
The single type-B vacancy features distinct $s$, $p$, and higher-energy gap-induced states ($\alpha$, $\beta$).
For the nearest-neighbor pair the bonding ($\sigma_{s}$) and antibonding ($\sigma^{*}_{s}$) hybrids of the $s$ orbital (Fig.~\ref{vacBB1_FDSTS}a,c), and a $p$ hybrid (Fig. S8a) are shown.
In the weakly interacting dimer, no $s-s$ interaction is discernible, while the $p-p$ interaction results in a slightly broadened $\sigma_{p}+\sigma^{*}_{p}$ feature. 
The insets show STM topographies of the corresponding vacancy configurations.
The upper axis indicates the orbitals' binding energy referenced to the onset of the 2D interface state of PTCDA/Ag(111).}
\label{summary_hybrid} 
\end{figure}

The coupling behaviour is summarised in Fig.~\ref{summary_hybrid}, showing \textit{local} EDHs (see Methods) of the artificial orbitals plotted on a linear scale. 
For the nearest-nighbor pair, the interaction of two $s$ orbitals gives rise to the expected bonding ($\sigma$) and antibonding ($\sigma^{*}$) hybrids.
Notably, the splitting is not symmetric with respect to the single-vacancy $s$ level: while $\sigma$ is stabilised by approximately 100~meV, $\sigma^{*}$ is shifted upward by about 106~meV.
Such asymmetric stabilisation and destabilisation is well known \cite{hoffmann_how_1987}.
Because the orbitals centred at different sites are non-orthogonal, the antibonding hybrid is more destabilising than the bonding hybrid is stabilising.
As the degree of coupling depends strongly on vacancy separation, the weakly coupled pair exhibits only a small energy shift.
Notably, the EDHs of the pair comprising bonding and antibonding hybrids shift to higher energies, consistent with the argument discussed above.
Together, these observations reinforce the close analogy between artificial and natural atoms, showing that the asymmetric stabilisation of bonding and antibonding hybrids, familiar from molecular chemistry, can also be observed in artificial-atom systems.

\subsection{High-energy orbitals induced by the corrugation of the 2D vacuum}

The artificial-atom orbitals discussed so far closely follow the conventional chemical analogy. 
This reflects the nearly parabolic dispersion of the 2D interface state near the $\bar\Gamma$ point (Fig.~\ref{vacB_STS}f), where the vacancy potential binds states resembling $s$ and $p$ orbitals of real atoms. 
At larger momenta, however, the electronic environment is no longer free-electron-like: the periodically corrugated potential of the PTCDA layer opens partial band gaps at the Brillouin-zone (BZ) boundaries \cite{Sabitova2018PRB}. 
Although the two lowest lying gaps are directional rather than complete, they appear clearly in the EDH of pristine PTCDA/Ag(111) (Fig.~\ref{vacB_STS}e).

In contrast, the EDH of a type-B vacancy reveals additional states, labelled $\alpha$ and $\beta$, emerging inside these band gaps. 
These resonances have no direct analogue in natural atoms. 
Despite their comparatively low spectroscopic weight and energetic overlap with the continuum density of states, real-space images of $\alpha$ and $\beta$ can be resolved in the experimental FDMs (Figs.~\ref{vacB_FDSTS_u}a,c). 
The corresponding TB FDMs (Figs.~\ref{vacB_FDSTS_u}b,d) reproduce the essential spatial features. 

\begin{figure}
\centering
\includegraphics[width=15cm]{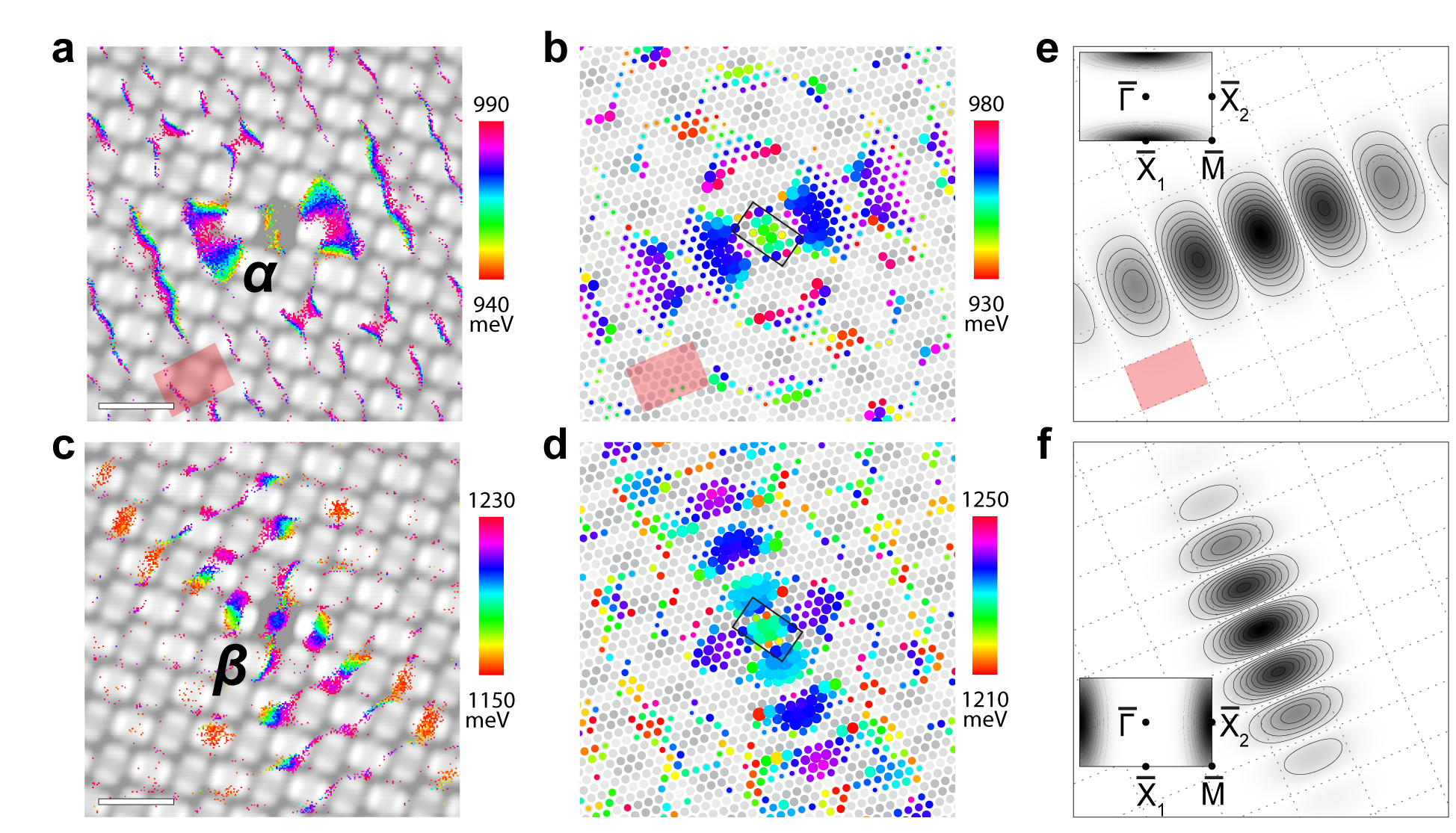}
\caption{\textbf{High-energy orbitals of a single vacancy.} 
(\textbf{a,c}) STM image of a type-B vacancy overlaid with a feature-detection map (FDM; see Methods) revealing $\alpha$ and $\beta$ states;  made from the corresponding portions of the peak-detection statistics marked with "$\mathrm{\alpha}$" and "$\mathrm{\beta}$" in Fig.~\ref{vacB_STS}e.
The colour reflects the peak energy, mapped according to the colorbar next to the image.
(\textbf{b,d}) Corresponding tight-binding (TB) simulated feature-detection maps (FDMs; see Methods).
The black rectangle indicates the model vacancy potential.
The colour reflects the peak energy, mapped according to the colorbar next to the image.
The size of the coloured nodes reflects the spectral function peak intensity. 
(\textbf{e,f}) Intensity plots of the two weighted plane-wave sums (see text) with the insets showing the weighting function plotted over the molecular lattice Brillouin zone (Fig.~\ref{vacB_STS}f).}
\label{vacB_FDSTS_u} 
\end{figure}

Although $\alpha$ and $\beta$ appear localised around the vacancy, their structure differs fundamentally from that of the low-energy $s$ and $p$ orbitals. 
As seen in Fig.~\ref{vacB_STS}f, both states emerge in the vicinity of band edges created by the BZ-boundary gaps, rather than splitting off from a parabolic band minimum. 
Their pronounced quasi-one-dimensional character is therefore inherited from the states of the surrounding electronic vacuum, i.e. the plane-wave components present at the edges of the corresponding band gaps \cite{Sabitova2018PRB}.

To make this connection explicit, we analyse their momentum composition. 
The $\alpha$ state is dominated by plane-wave components with $\mathbf{k}=\pm \bar X_1$, whereas $\beta$ is primarily composed of components with $\mathbf{k}=\pm \bar X_2$. 
By selecting plane waves within the molecular BZ and forming weighted sums (see Methods), we construct real-space intensities from contributions localised near $\pm \bar X_1$ and $\pm \bar X_2$ (insets of Figs.~\ref{vacB_FDSTS_u}e,f). 
The resulting patterns (Figs.~\ref{vacB_FDSTS_u}e,f) closely resemble the experimental and TB FDMs (Figs.~\ref{vacB_FDSTS_u}a-d), directly revealing how the $\alpha$ and $\beta$ states arise from the underlying band structure.

The gaps hosting these states are centred at the $\pm \bar X_1$ and $\pm \bar X_2$ points of the BZ. 
At each gap edge, the interface band forms standing waves \cite{Sabitova2018PRB} whose periodicities match the corresponding molecular lattice vectors, $\mathbf{D_1}$ and $\mathbf{D_2}$ (Fig.~\ref{vacB_STS}b). 
The vacancy selects and localises these band-edge standing waves, giving rise to anisotropic states whose dominant $\mathbf{k}$-vectors are confined to narrow regions of the BZ. 
As a result, the $\alpha$ and $\beta$ states exhibit quasi-one-dimensional modulations along directions set by the underlying lattice. 

The phase of this modulation provides further insight. 
The standing waves associated with the upper and lower edges of each gap are phase-shifted by $\pi$ \cite{Sabitova2018PRB}. 
Because the $\alpha$ state lies in close proximity to the upper gap edge (Fig.~\ref{vacB_STS}f), it is expected to be dominated by the corresponding standing wave. 
The good agreement between the modulation phase observed in the experimental and TB FDMs and the reconstructed pattern in Fig.~\ref{vacB_FDSTS_u}e confirms that $\alpha$ is predominantly a vacancy-pinned upper-gap band-edge state.

In contrast, the $\beta$ state lies closer to the middle of its gap (Fig.~\ref{vacB_STS}f), where contributions from both upper and lower band edges are energetically allowed. 
The more complex patterns observed in Figs.~\ref{vacB_FDSTS_u}c,d, therefore likely reflect a superposition of band-edge standing waves, and possibly the presence of multiple closely spaced states.

Taken together, these observations show that the high-energy vacancy states are largely dictated by the structure of the surrounding band dispersion. 
The vacancy acts primarily as a pinning centre that localises standing-wave patterns inherited from the band edges, underscoring the decisive role of the structured electronic vacuum in shaping the orbital spectrum of the artificial atom.

\section{Discussion}

Our results show that the orbital picture remains highly robust even in engineered low-dimensional systems. 
Artificial atoms patterned in a 2D electron vacuum display bound states that closely resemble the familiar $s$ and $p$ orbitals of natural atoms, and their interactions follow the same principles of spatial overlap and coupling that underpin chemical bonding.
In this picture, the strength of orbital coupling is primarily determined by three factors: the spatial separation between the artificial atoms, which controls the magnitude of orbital overlap; the symmetry of the interacting orbitals, which governs whether coupling is allowed or suppressed; and the energetic alignment of the orbitals with respect to the surrounding electronic vacuum, which sets their degree of localisation.

At the same time, key differences arise from the defining characteristics of the artificial environment.
First, the confining potential of a molecular vacancy is neither deep nor symmetric, which reduces orbital degeneracies and fixes preferred orientations, as seen in the single $p$ orbital vacancy state.
Second, the depth of the vacancy potential is inherently limited to a few hundred meV, implying that only a small number of bound states can form before merging with the continuum of the surrounding electronic vacuum.
This finite depth effectively restricts the "periodic table" of artificial atoms.
Still, it also provides an unusual degree of tunability: by locally modifying the potential landscape, one can design the number, symmetry, and energy spacing of the available orbitals.
Possibilities include deepening the potential by placing a donor in the vacancy, which, in its ionised form, attracts electrons \cite{Sierda2023Science}, or by making the vacancy larger. 

Beyond the analogues of conventional atomic orbitals, artificial atoms hosted in a gapped 2D vacuum develop additional bound states with no natural-atom counterparts. 
As shown in Fig.~\ref{vacB_FDSTS_u}, such states appear at higher energies within the partial band gaps of the electronic dispersion in vacuum. 
These gap-induced states can be viewed as orbitals formed in a structured electronic vacuum. 
Their existence demonstrates that the orbital concept continues to apply even when the underlying electronic vacuum has a strongly non-parabolic dispersion, and it extends it further: the allowed orbital spectrum is shaped not only by the local potential but also by the structure and gaps of the host band.

\section{Conclusions}

The higher-lying, in-gap orbitals identified in this work naturally connect to defect-induced electronic states in two-dimensional materials, where local potentials similarly give rise to fully- and quasi-localised states within band gaps \cite{bertoldo_quantum_2022, komsa_native_2015}. 
In contrast to the present metal-supported system, such defect states in semiconducting and insulating 2D materials often exhibit substantially longer lifetimes \cite{schuler_large_2019, qiu_atomic_2024}. 
Deliberately engineering and coupling such defect states can form designer electronic structures with emergent functionalities \cite{fang_atomically-precise_2023}. 
Viewed in this context, the gap-induced orbitals of artificial atoms reported here provide a conceptual bridge between defect physics in low-dimensional materials and the language of chemistry. 
By directly imaging these states in real space, our work demonstrates that the dispersion and gaps of the surrounding electronic vacuum determine the formation and shape of the orbitals, translating defect-induced electronic structure into an orbital language used in chemistry.

\section{Methods}

\subsection{Sample preparation and $dI/dV$ measurements}

The Ag(111) surface was cleaned by repeated cycles of Ar$^+$ sputtering and annealing to 420$^\circ$C under ultra-high vacuum. 
PTCDA molecules were thermally evaporated from a home-built Knudsen cell held at 298$^\circ$C onto the Ag(111) surface maintained at room temperature.
After deposition, the sample was briefly flashed to 200$^\circ$C, cooled to 90 K, and transferred into a CREATEC scanning tunnelling microscope (STM) operating at a base temperature of 6 K. 

Differential conductance $dI/dV$ spectra were recorded as $255\times255$-pixel grids over $20\times20$~nm$^2$ or $10\times10$~nm$^2$ surface areas, scanning the tip in constant-current mode with a set-point of $I = 100$~pA and $V = -300$~mV, using a lock-in modulation amplitude of $V_{\mathrm{mod}} = 20$~mV at $f_{\mathrm{mod}} = 6333$~Hz. 
Each spectrum was acquired in 2 s, resulting in a total grid-acquisition time of $\approx38$ h. 
All STM images were obtained under the same tunnelling conditions unless stated otherwise.

\subsection{Fitting with the bound state spectrum with the Newns--Anderson model}

We fitted the bound-state spectrum of the type-B vacancy in Fig. \ref{vacB_STS}d using the modified Newns-Anderson (NA) model developed in refs. \onlinecite {Limot2005PRL, Kroger2005PSS}.
This model treats the vacancy as an energy level, $\varepsilon_a$, interacting with both bulk Bloch states and the two-dimensional (2D) interface state of PTCDA/Ag(111). 
The fit is produced by calculating the density of states projected on the vacancy

\begin{equation}
n(E) = \frac{1}{\pi} \frac{\Delta(E)}{[E - \varepsilon_a - K(E)]^2 + \Delta(E)^2},
\end{equation}

\noindent where $K(E) + i\Delta(E)$ is a self energy. The  imaginary part of the self-energy is given by 

\begin{equation}
\Delta(E)=\Delta_b+\Delta_s \left[\frac{1}{2} + \frac{1}{\pi}\arctan \frac{2(E-E_0)}{\Gamma}\right],
\end{equation}

\noindent where $\Delta_b$ and $\Delta_s$ define the coupling strengths to the bulk and surface states, respectively, and the expression in brackets models the step-like density of state of the 2D interface state with the onset at $E_0$ and lifetime $\Gamma$. The real part of the self-energy $K(E)$ is

\begin{equation}
K(E) = \frac{\Delta_s}{2\pi} \ln \left[(E - E_0)^2 + \left(\frac{\Gamma}{2}\right)^2\right] + \text{const}.
\end{equation}

We fitted the spectrum in Fig.~\ref{vacB_STS}d using the experimental value  $E_0=710$ meV and the known value of $\Gamma=10$ meV \cite{Schwalb2008PRL}, leaving $\varepsilon_a, \Delta_b$, and $\Delta_s$ as free parameters. To reproduce the negative differential resistance observed in the experimental $dI/dV$ spectrum, we also included the energy dependence of the tunnelling-barrier transmission coefficient \cite{wagner_evaluation_2007}. Junction parameters were set to a tip-surface distance $d=12$ \AA~and the surface workfunction energy of $\Phi=4.5$ eV for both tip and surface. The fit shown in Fig. \ref{vacB_STS}d was obtained with $\Delta_b=77$ meV, $\Delta_s=485$ meV, and $\varepsilon_a=711$ meV.

\subsection{Feature detection scanning tunnelling spectroscopy (FD STS)}

Because the standard $dI/dV$ imaging fails (see Supplement), we had to use feature-detection scanning tunnelling spectroscopy (FD STS) \cite{Sabitova2018PRB, martinez-castro_disentangling_2022} to visualise the spatial distribution and energy evolution of vacancy-induced spectroscopic features. 
In this method, $dI/dV(V)$ spectra are recorded over a grid of points while scanning the area of interest in constant-current mode. 
Each spectrum is smoothed with a 120~meV window Savitzky-Golay filter to reduce experimental noise, and local maxima are identified as spectral peaks \cite{martinez-castro_disentangling_2022}. 
The bias voltages corresponding to these peaks are stored as "peak-detection" statistics.

To construct an energy-distribution histogram (EDH), the detected peak energies are aggregated into a histogram that shows how frequently each energy occurs. 
Because each grid pixel (i.e., spectrum) contributes at most one detection per energy, the EDH represents the fraction of grid pixels where a peak at that energy is detected. 
To ensure comparability between data sets, each EDH is normalised to the total number of grid pixels.
Note that because we only detect peaks, the step-like feature of the 2D interface state does not contribute to the EDH (Fig.~\ref{vacB_STS}e).

Selecting an energy interval within an EDH allows the reconstruction of a feature-distribution map (FDM) that marks all grid pixels whose spectra contain a peak within the chosen energy interval. 
The colour of each pixel in an FDM encodes the precise energy of the detected resonance. 
Consequently, FDMs reveal both the spatial location of $dI/dV$ features and their energy variation ("dispersion") across the surface.
This capability also makes FD STS useful for distinguishing scattering resonances from localised bound states. 
For instance, in the FDM shown in Fig.~\ref{vacB_FDSTS_p}a, the scattering states of the 2D interface state located further from the vacancy display strong energy dispersion that spans the full energy range of the FDM.

In contrast, the FDMs of localised orbitals ($s$ and $p$ states localised directly over the vacancy in Figs. \ref{vacB_FDSTS_s}a, \ref{vacB_FDSTS_p}a) exhibit a weaker dispersion. 
Although a stationary quantum state should have a fixed energy, the experimental FDMs suggest that these states shift by 10-30 meV depending on the STM tip position. This apparent variation in energy is attributed to experimental artefacts rather than intrinsic effects. 
Several factors may contribute:
(i) topographic height differences that modify the local voltage drop in the tunnelling junction;
(ii) changes in the tunnelling pathway, where over molecular sites the current may partially flow through broad molecular resonances, producing a Fano-like energy shift; and
(iii) weak coupling between the vacancy state and tip states. 
The latter effect has been observed recently but appears too small to fully account for the present observations \cite{stilp_very_2021}.
The energy dispersion seen at the very rims in the experimental as well as the theoretical FDMs (see e.g. Figs. \ref{vacB_FDSTS_s}a,b) is due to the artefact introduced by the peak detection algorithm: Towards the edges, the peak intensity always drops, making its accurate detection more difficult.
In particular, the energy of a weak peak may be shifted by another, more intense feature whose energy is nearby. 
Because of the same reason, the considerable energy dispersion of the $\alpha$ and $\beta$ orbitals in experimental FDMs (Figs. \ref{vacB_FDSTS_u}a,c) is primarily due to their small intensity.

Finally, FD STS allows the preparation of EDHs obtained from the areas of the $dI/dV$ grids where a particular localised state has been observed in an FDM. These \textit{local} EDHs, shown in Fig.\ref{summary_hybrid}, thus filter out the statistics that do not contribute to the given feature, facilitating its individual analysis.

\subsection{Tight binding simulations}

Electronic states associated with PTCDA/Ag(111) vacancies were modelled using a two-dimensional tight-binding (TB) approach on a $500\times 500$-node lattice with open boundary conditions (see Supplement)..
Each node represents an atomic site of the outermost Ag(111) layer, with coordinates $\mathbf{r}_i$. 
Only nearest-neighbour hopping was included, with the hopping integral chosen to reproduce the parabolic dispersion of the PTCDA/Ag(111) interface state corresponding to an effective mass of $m_\mathrm{eff}=0.47\,m_\mathrm{e}$ [\onlinecite{Temirov2006Nature}]
The on-site potential of each node, $U(\mathbf{r}_i)$, was offset by $U_0=710$ meV to match the onset energy of the interface state (see Figs.~\ref{vacB_STS}d,f).

The PTCDA monolayer forms a commensurate monolayer with the lattice vectors $|\mathbf{D_1}| =18.9$~\AA, and $|\mathbf{D_2}| =18.9\times 12.6$ \AA~[\onlinecite{Glockler1998SS}] and an almost rectangular unit cell as shown in Figs.~\ref{vacB_STS}b.
Because the adsorption sites of both A and B molecules are known \cite{Rohlfing2007}, the registry between the PTCDA layer and the TB lattice can be fixed, allowing us to impose the periodic molecular potential $U_\mathrm{PTCDA}(\mathbf{r}_i)$ at each atom position $\mathbf{r}_i$.
Following our previous work \cite{Sabitova2018PRB}, this periodic potential was expressed as

\begin{equation} 
U_{\mathrm{PTCDA}}(\mathbf{r}_i) = \sum_\mathbf{P}U_\mathbf{P} e^{i\mathbf{P}\cdot\mathbf{r}_i}\,, 
\label{perU}
\end{equation}

\noindent where the eight Fourier components correspond to the combinations of vectors $\mathbf{B}_1$ and $\mathbf{B}_2$ reciprocal to $\mathbf{D}_1$ and $\mathbf{D}_2$  (see Supplement) with amplitudes $U_{\pm\mathbf{B}_1}=38$ meV, $U_{\pm\mathbf{B}_2}=57$ meV, and $U_{\pm(\mathbf{B}_1\pm\mathbf{B}_2)}=-61$ meV [\onlinecite{Sabitova2018PRB}].
The total potential at each site was thus $U(\mathbf{r}_i)=U_0+U_{\mathrm{PTCDA}}(\mathbf{r}_i)$.

To introduce a single type-B molecular vacancy, we selected a corresponding adsorption site near the centre of the TB cluster to avoid boundary effects.
The vacancy potential was modelled by defining a rectangular region, $1.4\times$ the size of a PTCDA molecule ($11.5 \times 6.8$ \AA), centred at the chosen site with its molecular axis oriented according to the literature \cite{Glockler1998SS}.
The on-site potential within this region was modified according to 

\begin{equation} 
U'(\mathbf{r}_i)=U(\mathbf{r}_i)\times\left[\frac{1}{2}+\frac{1}{\pi}\arctan{\frac{f(\mathbf{r}_i)}{s}}\right],
\label{vacU}
\end{equation}

\noindent where $f(\mathbf{r}_i)$ is the signed distance from the rectangle boundary and $s=0.3$ is a smoothing parameter.
The resulting potential profiles are shown in the Supplement.
To reproduce the experimentally observed orientation of the $p$ orbital in Fig.\ref{vacB_FDSTS_p}a, the vacancy rectangle had to be additionally rotated clockwise by 10$^\circ$ relative to the long molecular axis (Fig.\ref{vacB_FDSTS_p}b).
All TB simulations of the vacancy dimers used these "rotated" vacancy potentials.

Spectral functions were computed for all atoms within a circular area of radius $50 \times$ TB-lattice nearest-neighbour distance around the vacancy site using the Lanczos algorithm \cite{lanczos_iteration_1950} (see Supplement), providing quantities directly comparable to the local density of states measured by STM.
FD STS analysis was then applied to the calculated spectral functions to obtain theoretical feature-distribution maps (TB FDMs) shown in Figs.~\ref{vacB_FDSTS_s}b, \ref{vacB_FDSTS_p}b, \ref{vacBB1_FDSTS}b,d,f,h and \ref{vacB_FDSTS_u}b,d.
The TB-calculated spectral functions provide clear intensity information, free of parasitic crosstalk with the vacancy topography present in the experimental $dI/dV$ data (see above and Supplement).   
We therefore exhibit this information as the size of the corresponding lattice node in the TB FDMs.
For comparison, the TB FDMs that omit the intensity information are shown in the Supplement.

\subsection{Differences between experimental and calculated feature distribution maps (FDMs).}

All orbital energies of the simulated bound states come out lower than in the experiment, suggesting that the real vacancy potential is shallower than we assumed.
The simulated bound states exhibit no colour dispersion, which confirms our assumption that the dispersion observed in the experiment is a measurement artefact (see above). 
The substantially larger size of the simulated bound states highlights the essential difference between the two setups: In the experiment, bound-state resonances broaden due to their coupling to the bulk (see above), making detection difficult as the peak intensity decreases. 
This leads to a reduction in the area of the experimental FDM where the bound-state peak was detected.
Because the simulated system is free of the lifetime effects and noise, the corresponding resonance is detected more efficiently, thereby extending its detection area in the theoretical FDM.

\subsection{Simulating $\alpha$ and $\beta$ with plane waves}

To analyse the $\mathbf{k}$-space composition of the $\alpha$ and $\beta$ states, we constructed weighted superpositions of plane waves of the form
\[
\sum_{\mathbf{k}} A(\mathbf{k}) e^{i(\mathbf{k}\cdot\mathbf{r} + \phi)},
\]
where the weighting function
\[
A(\mathbf{k}) = A(k_x,k_y) = \exp\left[-\frac{1}{2}\left(\frac{k_x-a_x}{\sigma_x}\right)^2\right] \exp\left[-\frac{1}{2}\left(\frac{k_y-a_y}{\sigma_y}\right)^2\right]
\]
is a Gaussian in $\mathbf{k}$-space with parameters listed in Table~\ref{table1}.
The phase $\phi=\pi$ was chosen to reproduce the FDMs shown in Figs.~\ref{vacB_FDSTS_u}a--d. 
This phase corresponds to the standing-wave solutions at the upper edges of the respective band gaps \cite{Sabitova2018PRB}.

\begin{table}[htbp]
\centering
\caption{Parameters used to define $k$ distributions for modelling $\alpha$ and $\beta$ states.}
\label{tab:alpha_beta_parameters}
\begin{tabular}{l|c|c}
   & $\alpha$, $[\pi/|\mathbf{D_1}|]$ & $\beta$, $[\pi/|\mathbf{D_2}|]$\\ 
  \hline
  $(a_x,\sigma_x)$ 
    & $(1,0.2)$ 
    & $(0,0.5)$ \\
  $(a_y,\sigma_y)$ 
    & $(0,0.8)$ 
    & $(1,0.2)$ \\
\end{tabular}
\label{table1}
\end{table}

\section*{Acknowledgements}

R.T. thanks Jeff Rawson for fruitful discussions. 

\section*{Author Contributions}

F.S.T. and R.T. conceived the research.
M-W.G., A.S., T.E. and R.T.  conducted the experiments. 
M-W.G. and R.T. analysed the experimental data with input from T.E. and C.W.
A.R. wrote the code for TB simulations. 
R.T. and A.R. performed the simulations. 
F.S.T. and R.T. wrote the paper with significant contributions from all authors.  

\section*{Funding}
Financial support by the Deutsche Forschungsgemeinschaft through SFB 1083 (project-ID 223848855) is acknowledged.
 
\section*{Competing Interests}
The authors declare no competing interests.

\clearpage
\appendix
\setcounter{section}{0}
\renewcommand{\thesection}{S\arabic{section}}
\setcounter{figure}{0}
\renewcommand{\thefigure}{S\arabic{figure}}
\setcounter{equation}{0}
\renewcommand{\theequation}{S\arabic{equation}}
\makeatletter
\renewcommand\appendixname{}
\makeatother

\section*{Supplementary Information}

\section{$dI/dV$ imaging of the vacancy bound states}

\begin{figure}[!h]
\centering
\includegraphics[width=16cm]{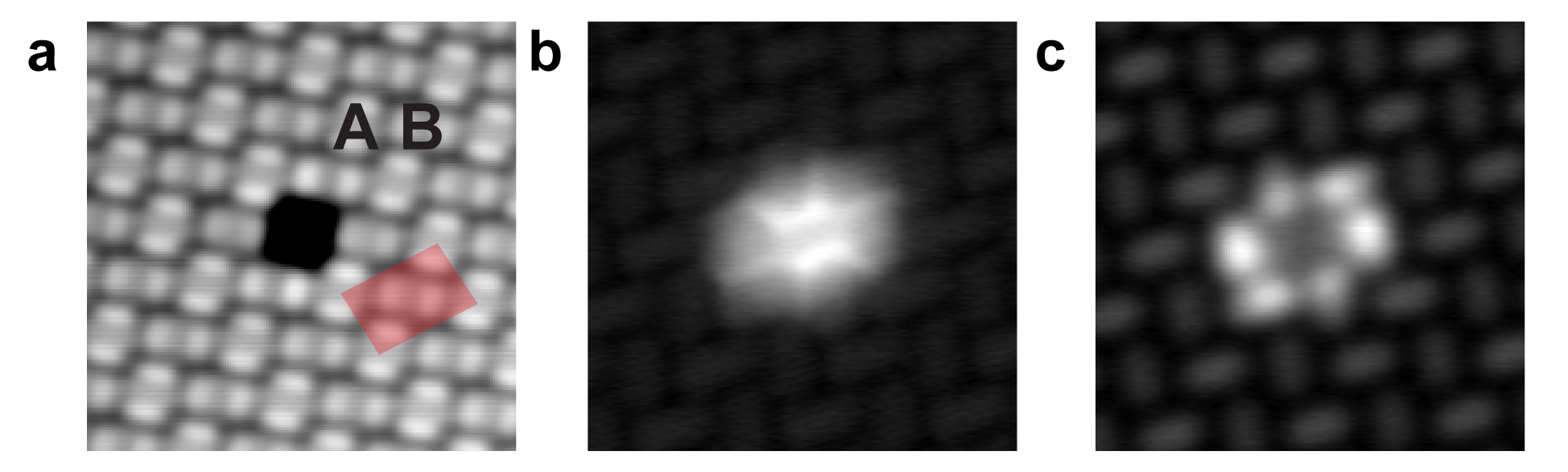}
\caption{(\textbf{a}) A fragment of the PTCDA/Ag(111) monolayer with a single type-A defect. (\textbf{b}) Constant current $dI/dV$ image of the area from (a) taken at the energy of the $s$ orbital (see main text).(\textbf{c}) Constant height $dI/dV$ image of the area from (a) taken at the energy of the $s$ orbital (see main text).}
\label{SF_FDSTS} 
\end{figure}

Fig.~\ref{SF_FDSTS} exposes the problems that arise in the process of visualizing the vacancy bound states when using traditional $dI/dV$ imaging approaches. 
Neither the constant-current, nor the constant-height $dI/dV$ imaging capture the bound state correctly because each of them plots the intensity of the $dI/dV$ signal that is affected by the changing topography of the sample. 
In contrast, FD STS discards the $dI/dV$ intensity by outputting only the event of a peak detection (see main text), which makes its output free of the effects of the topography.

\section{Details of Tight-binding simulations}

\subsection{Tight-binding Hamiltonian}

To describe the 2D interface state of PTCDA/Ag(111), we use a single-band model.
Although one typically uses a nearly-free-electron description with an effective electron mass for this purpose, in the present case, it is more convenient to set up an effective tight-binding formulation.
The nearest-neighbour tight-binding Hamiltonian for a triangular lattice describing the (111) surface is given by

\begin{align}
    \hat{\mathcal{H}}
    = 
    -t\sum_{\langle jk\rangle} c_j^\dagger c_k
    =
    -t\sum_\mathbf{r}\sum_{i = 1}^6c^\dagger_\mathbf{r}c_{\mathbf{r}+\mathbf{l}_i}
    \,,
\end{align}

where $-t$ is the hopping parameter and $\mathbf{l}_i$ are vectors pointing to the six nearest neighbors.
Taking the lattice vectors as $\mathbf{d}_1 = d\hat{x}$ and $\mathbf{d}_2=d \hat{x}/ 2 + d\sqrt{3}\hat{y} / 2$ (see unit cell depicted in the inset of Fig.2b in the main text), we get $\mathbf{l}_{1,2} = \pm \mathbf{d}_1$, $\mathbf{l}_{3,4} =\pm \mathbf{d}_2$, $\mathbf{l}_{5,6} = \pm (\mathbf{d}_1 - \mathbf{d}_2)$.
Next, we write $c_\mathbf{r} = N^{-1/2}\sum_\mathbf{q}c_\mathbf{q}e^{i\mathbf{q}\cdot\mathbf{r}}$, where $N$ is the number of unit cells, leading to

\begin{align}
    \hat{\mathcal{H}}
    &=
    -t\sum_\mathbf{r}\sum_{i = 1}^6
N^{-1}\sum_{\mathbf{qq}'}c_\mathbf{q}^\dagger e^{-i\mathbf{q}\cdot\mathbf{r}}
c_\mathbf{q'}e^{i\mathbf{q}'\cdot(\mathbf{r}+\mathbf{l}_i)}
    \nonumber
    \\
    &=
    -t\sum_{i = 1}^6
 \sum_{\mathbf{q}}c_\mathbf{q}^\dagger 
c_\mathbf{q}e^{i\mathbf{q}\cdot \mathbf{l}_i}
\nonumber
\\
&=
-2t\sum_\mathbf{q}
\left\{
\cos(\mathbf{q}\cdot\mathbf{d}_1)
+
\cos(\mathbf{q}\cdot\mathbf{d}_2)
+
\cos[\mathbf{q}\cdot(\mathbf{d}_1-\mathbf{d}_2)]\right\}c_\mathbf{q}^\dagger 
c_\mathbf{q}
\nonumber
\\
&=
-2t\sum_\mathbf{q}
\left[
\cos(q_x d)
+
2\cos(q_x d / 2)
\cos(\sqrt{3}q_y d / 2)\right]c_\mathbf{q}^\dagger 
c_\mathbf{q}\,.
\end{align}
To determine $t$, we expand $\hat{\mathcal{H}}$ for small $\mathbf{q}$

\begin{align}
    \hat{\mathcal{H}}
    &\approx
-2t\sum_\mathbf{q}\left[3 - \frac{3}{4}d^2q^2\right]c_\mathbf{q}^\dagger 
c_\mathbf{q}
=\sum_\mathbf{q}\left(-6t + \frac{3t d^2 q^2}{2}\right)c_\mathbf{q}^\dagger 
c_\mathbf{q}\,,
\label{eqn:Low-Energy-H}
\end{align}
so that $t = \hbar^2 / (3d^2m_* m_e)$, where $m_*$ is the effective electronic mass in units of the electronic mass $m_e$, making it possible to obtain $t$ from the band curvature.
Expressing $d$ in terms of Bohr radii $a_0$,  $t = 2\hbar^2 / (2\times3d^2a_0^2m_* m_e) = 2\mathrm{Ry}/(3d^2m_*) = \mathrm{Ha}/(3d^2m_*)$.

\begin{figure*}
    \centering
    \includegraphics[width = 16cm]{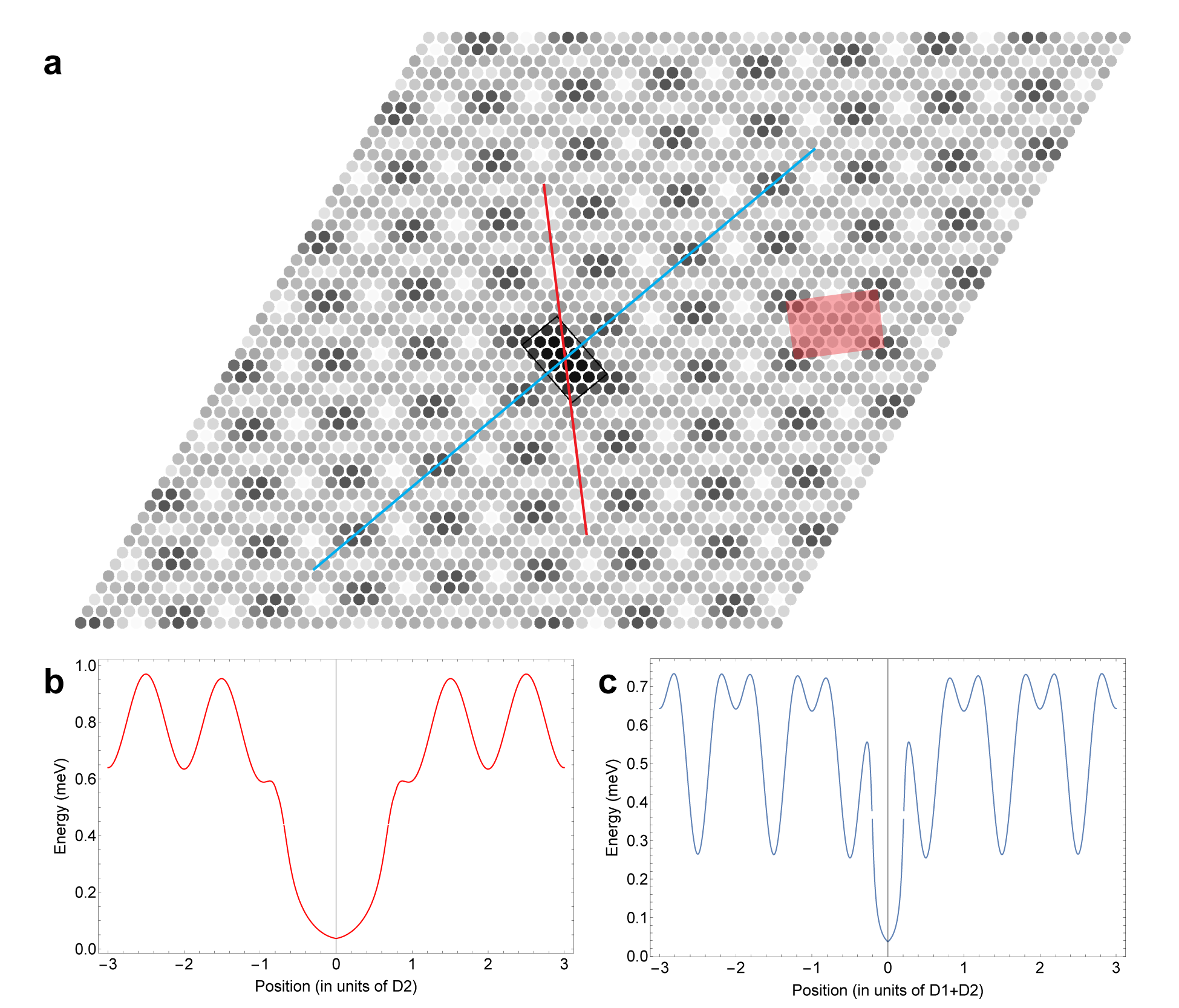}
    \caption{(a) Ag(111) lattice with the local potential at each site denoted by a grey colour (dark is more attractive). The red dashed rectangle labels the PTCDA unit cell, as shown in Fig. 1 of the main text. Red and blue lines label the locations of the potential profiles shown in Figs.~\ref{fig:SF_VAC}b-c. A black rectangle has been used to generate the potential of the B-vacancy (see Methods section of the main text). (b-c) Potential profiles plotted along the lines shown in Fig.~\ref{fig:SF_VAC}a}
   \label{fig:SF_VAC}
\end{figure*}

\subsection{Periodic potential}
\label{sec:Periodic_potential}

Tiling the silver surface with a molecular monolayer produces a periodic structure commensurate with the Ag(111) lattice, spanned by lattice vectors $\mathbf{D}_1 = 6\mathbf{d}_1 + \mathbf{d}_2$ and $\mathbf{D}_2 = -3\mathbf{d}_1 + 5\mathbf{d}_2$ (see the molecular unit cell in Fig.1b in the main text).
The simplest way to capture the effects of this monolayer is to assume that it generates a periodic potential so that the surface-state Hamiltonian acquires a position-dependent term

\begin{align}
    \hat{\mathcal{H}} &= 
    \sum_{\mathbf{q}} c^\dagger_\mathbf{q} \varepsilon_{\mathbf{q}} 
    c_{\mathbf{q}}
    +
    \sum_\mathbf{r} c_{\mathbf{r}}^\dagger U_\mathbf{r}c_{\mathbf{r}}
  \,.
    \label{eqn:Hamiltonian_periodic}
\end{align}

Writing the position of Ag atoms as $\mathbf{r}= m \mathbf{d}_1 + n \mathbf{d}_2$, the potential expression becomes

\begin{equation}
    U_\mathbf{r} = \sum_\mathbf{P}U_\mathbf{P}e^{i\mathbf{P}\cdot\mathbf{r}}e^{-i\mathbf{P}\cdot\mathbf{s}}\,,
    \label{eqn:U}
\end{equation}
where $U_\mathbf{P}$ are real and $\mathbf{P}$ belong to the potential's reciprocal lattice spanned by 

\begin{equation}
    \mathbf{B}_1 = 2\pi \frac{\mathbf{D}_2\times\hat{z}}{\mathbf{D}_1\cdot (\mathbf{D}_2\times\hat{z})}\,,
    \quad
    \mathbf{B}_2 = 2\pi \frac{\mathbf{D}_1\times\hat{z}}{\mathbf{D}_2\cdot (\mathbf{D}_1\times\hat{z})}\,,
\end{equation}
Including the vector $\mathbf{s}$ in Eq.~\eqref{eqn:U} enables us to "slide" the potential over the surface and, by choosing $\mathbf{s} = (\mathbf{d}_1 - \mathbf{d}_2) / 2$, we can use the data from ref.~[\onlinecite{Sabitova2018PRB}] (see main text) to get the potential at each atom site, as shown in Fig.~\ref{fig:SF_VAC}(a).

Rewriting $c_\mathbf{r}$ in terms of $c_\mathbf{q}$ in  Eq.~\eqref{eqn:Hamiltonian_periodic} yields

\begin{align}
    \hat{\mathcal{H}} &= 
    \sum_{\mathbf{q}} c^\dagger_\mathbf{q} \varepsilon_{\mathbf{q}} 
    c_{\mathbf{q}}
    +
    N^{-1}\sum_\mathbf{P}\sum_\mathbf{r} \sum_{\mathbf{qq}'}e^{-i\mathbf{q}'\cdot\mathbf{r}}c_{\mathbf{q}'}^\dagger U_\mathbf{P}e^{i\mathbf{P}\cdot\mathbf{r}}e^{-i\mathbf{P}\cdot\mathbf{s}}e^{i\mathbf{q}\cdot\mathbf{r}}c_\mathbf{q}
    \nonumber
    \\
    &= 
    \sum_{\mathbf{q}} c^\dagger_\mathbf{q} \varepsilon_{\mathbf{q}} 
    c_{\mathbf{q}}
    +
    \sum_\mathbf{P} e^{-i\mathbf{P}\cdot\mathbf{s}}\sum_{\mathbf{q}} c_{\mathbf{P}+\mathbf{q}}^\dagger U_\mathbf{P} c_{\mathbf{q}}
  \,.
    \label{eqn:Hamiltonian_periodic_momentum}
\end{align}
\begin{figure}
    \centering
    \includegraphics[width = 3.75in]{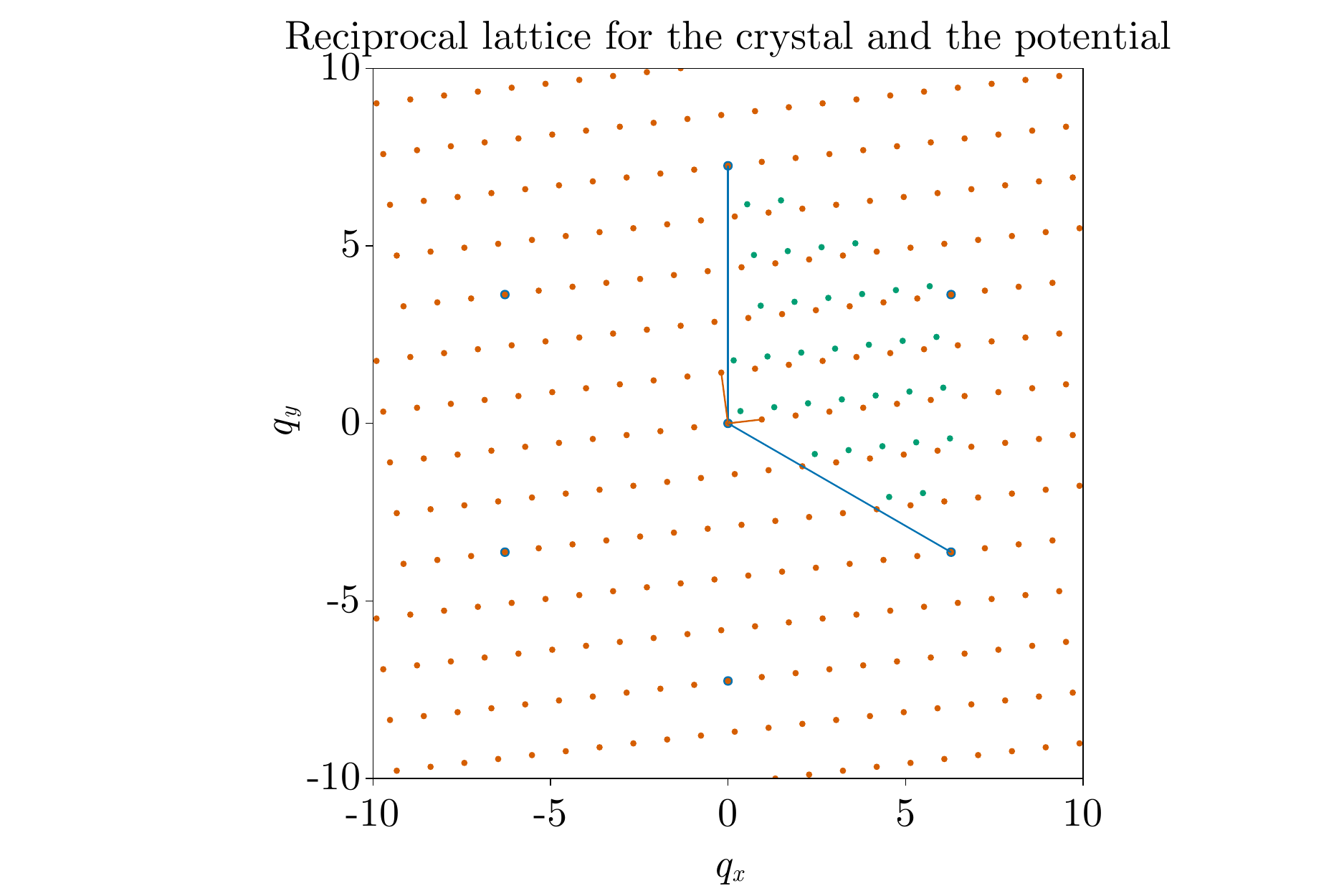}
    \\
    \includegraphics[width = 2.5in]{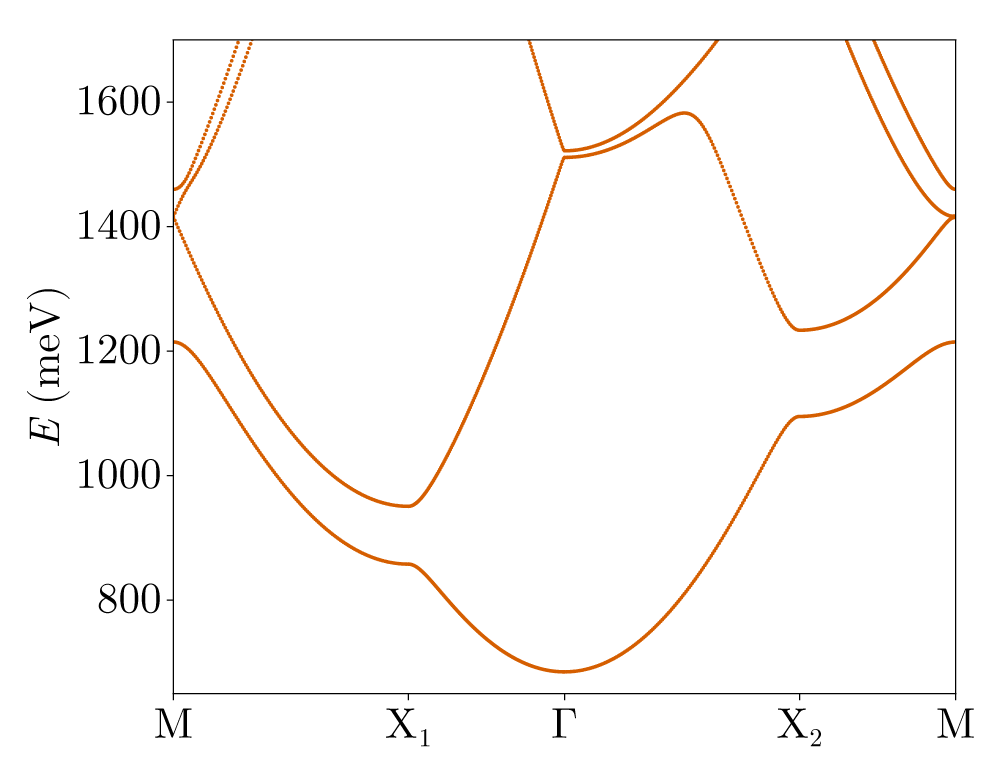}
    \caption{(Top) Reciprocal lattices for the crystal and the PTCDA monolayer-induced potential. The green dots are an example of a set of momenta within the unit cell coupled by the potential, in accordance with Eq.~\eqref{eqn:Hamiltonian_periodic_momentum}. (Bottom) Band structure produced by the molecular tiling of Ag(111) surface. We note that this band structure reproduces the one reported in ref.~[\onlinecite{Sabitova2018PRB}].}
    \label{fig:Lattice}
\end{figure}

Equation~\eqref{eqn:Hamiltonian_periodic_momentum} shows that the periodic potential couples the surface plane wave states whose momentum differs by a vector of the potential's reciprocal lattice.
Because of the commensurability between the molecular and silver lattices, their reciprocal lattices are also commensurate, as can be seen in Fig.~\ref{fig:Lattice}.
Additionally, since the real-space potential unit cell is 33 times larger than the silver unit cell, its reciprocal unit cell is 33 times smaller.
Consequently, any state with momentum $\mathbf{q}$ inside the first Brillouin zone couples to 32 other states inside the Brillouin zone, separated from it by momentum $\mathbf{P}$.
If $\mathbf{q} + \mathbf{P}$ lies outside the Brillouin zone, following Bloch's theorem, the momentum is transported back inside it by adding an appropriate reciprocal lattice vector $\mathbf{G}$.
Thus, the presence of the periodic potential generates $33\times 33$ Hamiltonian blocks of coupled states, where there is a single state inside the potential's unit cell.
This folding of the original silver Brillouin zone reduces its size, but generates multiple bands.
Conveniently, the potential-induced coupling between the states does not depend on $\mathbf{q}$, as can be seen from Eq.~\eqref{eqn:Hamiltonian_periodic_momentum}, and needs to be calculated only once.

The band structure calculated using Hamiltonian block diagonalisation for the folded Brillouin zone is shown in Fig.~\ref{fig:Lattice}.
The bottom of the band structure remains parabolic, while the potential-induced folding of the Brillouin zone leads to level repulsion, generating multiple bands.
Adding localised defects by "removing" individual adsorbed molecules breaks system periodicity, giving rise to oscillations in charge density.
However, unlike Friedel oscillations, which are traditionally discussed in the context of localised defects scattering electrons with parabolic dispersion, here the scattered electrons reside in a non-trivial vacuum generated by the periodic potential.

\subsection{Lanczos algorithm}

There are several ways to calculate the local density of states of introduced defects.
First, it is, in principle, possible to create a supercell that encloses the defects and is sufficiently large to avoid interaction effects from the neighbouring supercells.
Unfortunately, given the scale of the region studied experimentally, such supercells can be prohibitively expensive.
Another way involves calculating Green's functions propagators, treating the molecular vacancies as scattering potentials.
This strategy can also be costly because it requires multiple propagator calculations, each involving repeated integration over the Brillouin zone.
Therefore, we take a different approach and generate a finite system.
The size of the system, taken to be $500\times 500$ basis vectors (atoms), is chosen to be large enough to suppress spurious edge effects.
We then apply on-site potentials to each atom using the periodic form produced by the molecular layer.
The potential of a molecular vacancy is imposed according to the recipe discussed in the main text.
The resulting potential of a single type-B vacancy can be seen in Fig.~\ref{fig:SF_VAC}.

With 250,000 atoms, the Hamiltonian matrix would generally be unmanageable.
However, because each silver atom couples only to its nearest neighbours, the matrix is very sparse, making it tractable.
At the same time, matrix inversion required to calculate the spectral function cannot be performed directly because the result is generally not sparse.
Fortunately, we are only interested in the diagonal elements of the matrix inverse (corresponding to the spectral function), which allows us to utilise the Lanczos algorithm.
The Lanczos algorithm is an iterative method for approximating the diagonal elements of matrix inverses.~\cite{lanczos_iteration_1950}
In our case, we calculate the spectral function, given by the diagonal elements of $-\frac{1}{\pi}\left(E - H+i0\right)^{-1}$, where $E$ is the energy at which the spectral function is evaluated and $H$ is the real-space sparse hopping Hamiltonian (Eq.~\ref{eqn:Hamiltonian_periodic_momentum}).
Each diagonal element $-\frac{1}{\pi}\left(E - H+i0\right)^{-1}_{jj}$ gives the spectral function of the $j$th atom.

Following the algorithm, the diagonal elements are given by a continued fraction

\begin{align}
    \left(E - H+i0\right)_{jj}^{-1}&\approx\frac{1}{E-a_0-g_1}\,,
    \nonumber
    \\
    g_n &=\frac{b_n^2}{E - a_n - g_{n+1}}\,,
    \nonumber
    \\
    b_n &= \sqrt{\langle F_n|F_n\rangle }\,,
    \label{eqn:Lanscoz}
\end{align}
where the relevant terms are

\begin{align}
    |F_n\rangle &= H|f_{n-1}\rangle - a_{n-1}|f_{n-1}\rangle-b_{n-1}|f_{n-2}\rangle\,,
    \nonumber
    \\
    b_n &= \sqrt{\langle F_n|F_n\rangle }\,,
    \nonumber
    \\
    |f_n\rangle &=|F_n\rangle / b_n\,,
    \nonumber
    \\
    a_n &=\langle f_n |H|f_n\rangle\,.
\end{align}

The starting values are $a_0 = H_{jj}$, $b_0 = 0$, $|f_0\rangle$ is a one-hot at $j$ vector.
We can see that for a particular $j$, we need to calculate a set of $a_n$ and $b_n$ only once and then compute the approximate spectral function by inserting the desired value of $E$ into Eq.~\ref{eqn:Lanscoz}.

When choosing the number of steps in the recursion, it is necessary to balance accuracy and runtime.
We noticed that 1000 steps are sufficient to obtain converging results.
By repeating the procedure for each atom of interest, we obtain a collection of $a_n's$ and $b_n$'s, allowing us to generate spatially-resolved spectral maps.

\section{TB results for a single type-B vacancy}

\begin{figure}
\centering
\includegraphics[width=8cm]{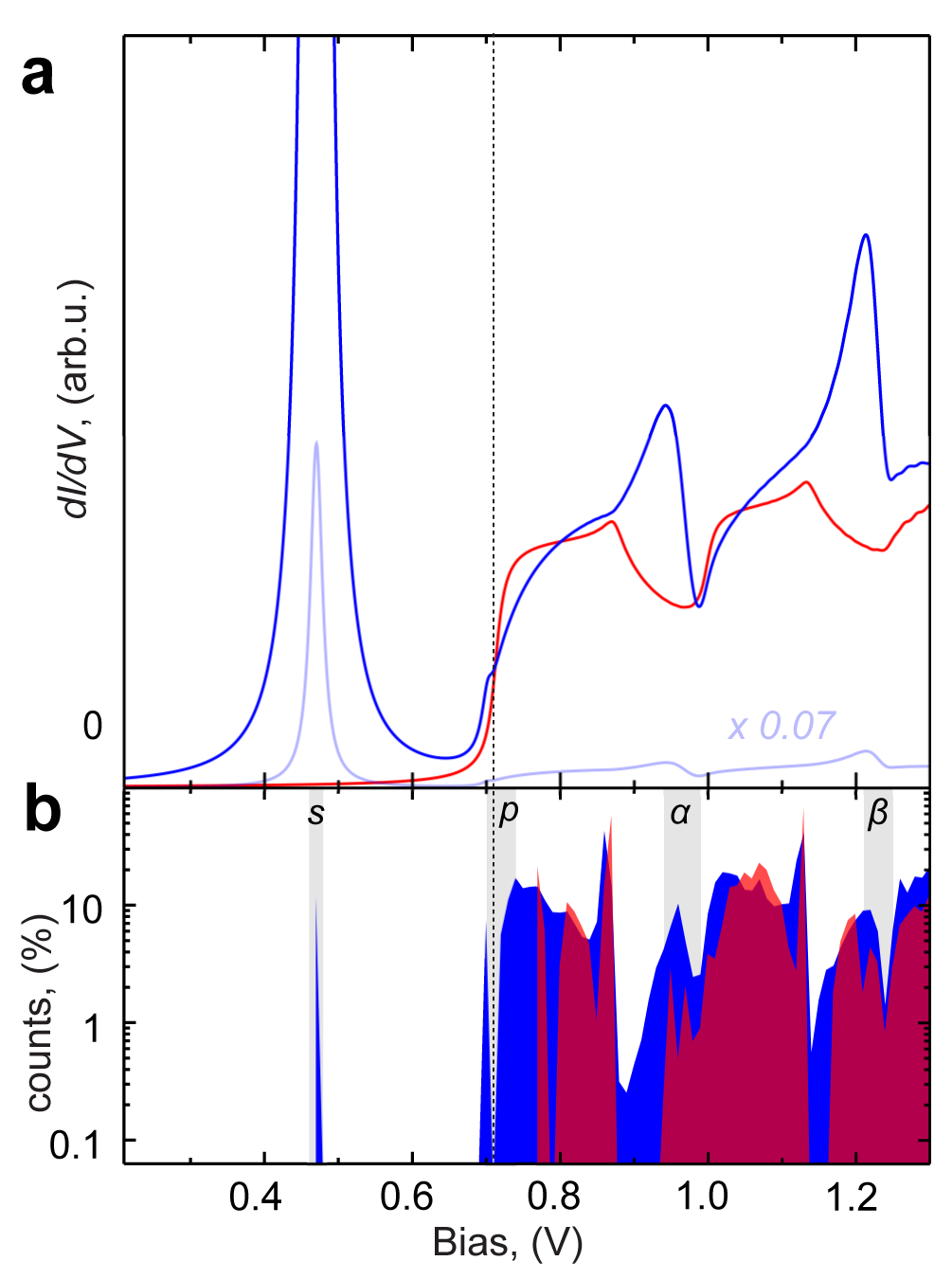}
\caption{(\textbf{a}) Spectral functions of individual atoms in the model cluster obtained by TB calculations with the potential of a single type-B vacancy shown in Fig.~\ref{fig:SF_VAC}. 
The blue spectrum was obtained on the TB lattice node nearest to the vacancy centre. 
The red spectrum was obtained on a node far from the vacancy and therefore exhibits an undisturbed spectrum of the model 2D interface state. (\textbf{b}) 
Two logarithmic energy distribution histograms (EDHs; see Methods) extracted by applying FD STS to the TB simulated data (see Methods section in the main text). 
The blue (red) EDH was obtained from TB data calculated on a cluster, with (without) the vacancy potential. 
The discrepancies between the EDHs indicate the presence of new vacancy-induced states. 
The black dotted line marks the onset of the 2D IS at 710 meV. 
Grey rectangles mark the energy intervals used to extract images of the bound states shown in the main text Fig.~2b~($s$), Fig.~3b~($p$), Fig.~6b~($\alpha$), Fig.~6d~($\beta$).}
\label{SF_EDH} 
\end{figure}

\newpage

\section{Simulated feature distribution maps of the vacancy orbitals without intensity}

Here we show analogues of Figs. 2,3,4,6 of the main text, with the difference that the TB FDMs used here do not show the intensity information (see Methods), thus becoming the full analogues of the experimental FDMs (see Methods).

\begin{figure}
\centering
\includegraphics[width=8cm]{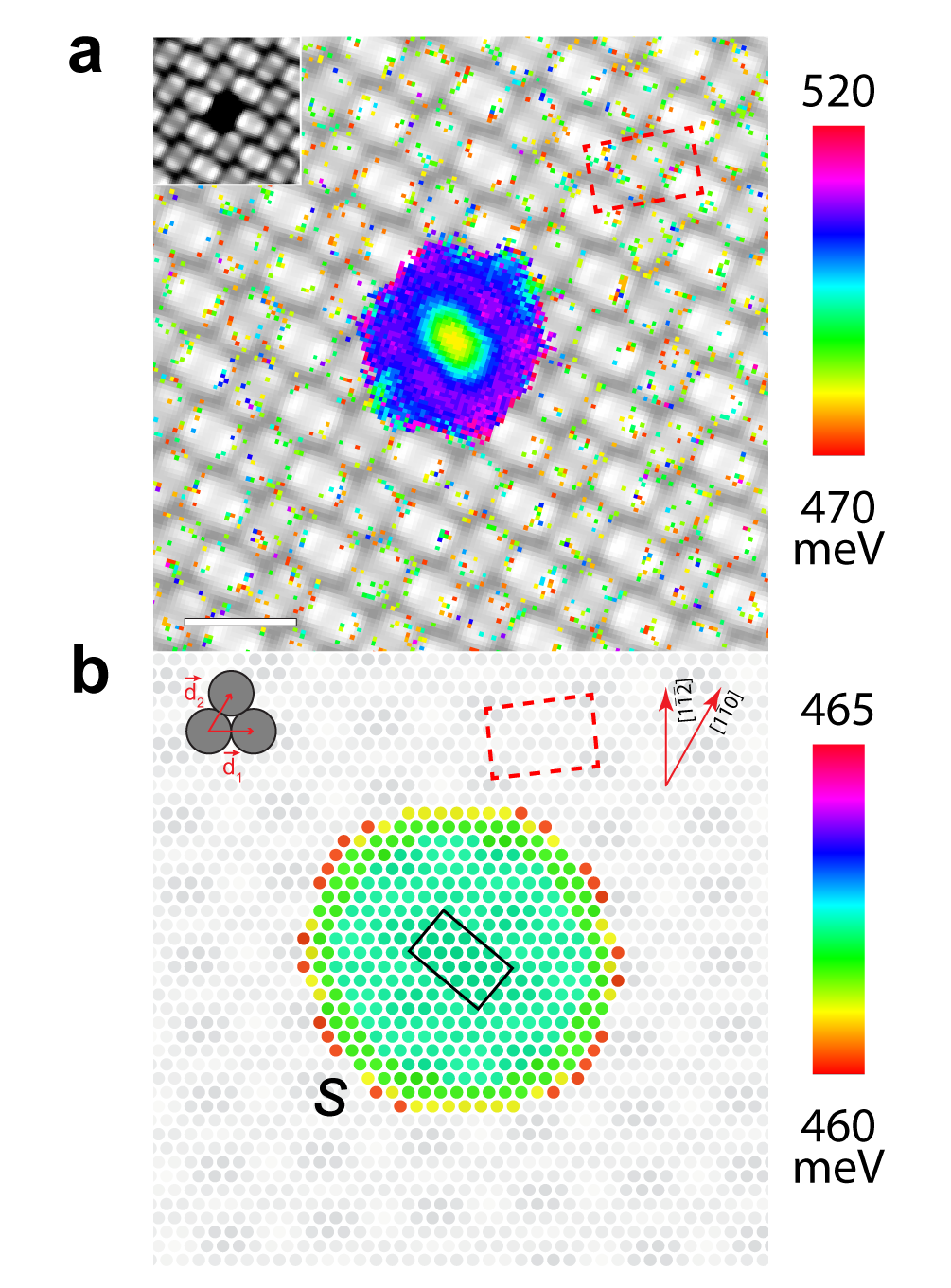}
\caption{Artificial $s$ orbital of a single molecular vacancy. 
(\textbf{a}) STM image of a type-B vacancy overlaid with a feature-detection map (FDM; see Methods) revealing the lowest-energy bound state of the vacancy; made from the portion of the peak-detection statistics marked with "s" in Fig.1e.
The colour reflects the peak energy, mapped according to the colorbar next to the image.
The inset provides a zoomed-out view of the vacancy, which in the main panel is covered by the FDM overlay.
(\textbf{b}) FDM obtained from the tight-binding (TB) simulation (see Methods).
The grey corrugation is due to the periodic potential (dark is more attractive).
The black rectangle was used to impose the vacancy potential.
The inset depicts the unit cell of the simulation lattice.
The colour reflects the peak energy, mapped according to the colorbar next to the image.
Red arrows indicate the high-symmetry directions of Ag(111), and the inset depicts the atomic lattice with the basis vectors $\mathbf{d_1}$ and $\mathbf{d_2}$.}
\label{SF_EDH} 
\end{figure}

\begin{figure}
\centering
\includegraphics[width=8cm]{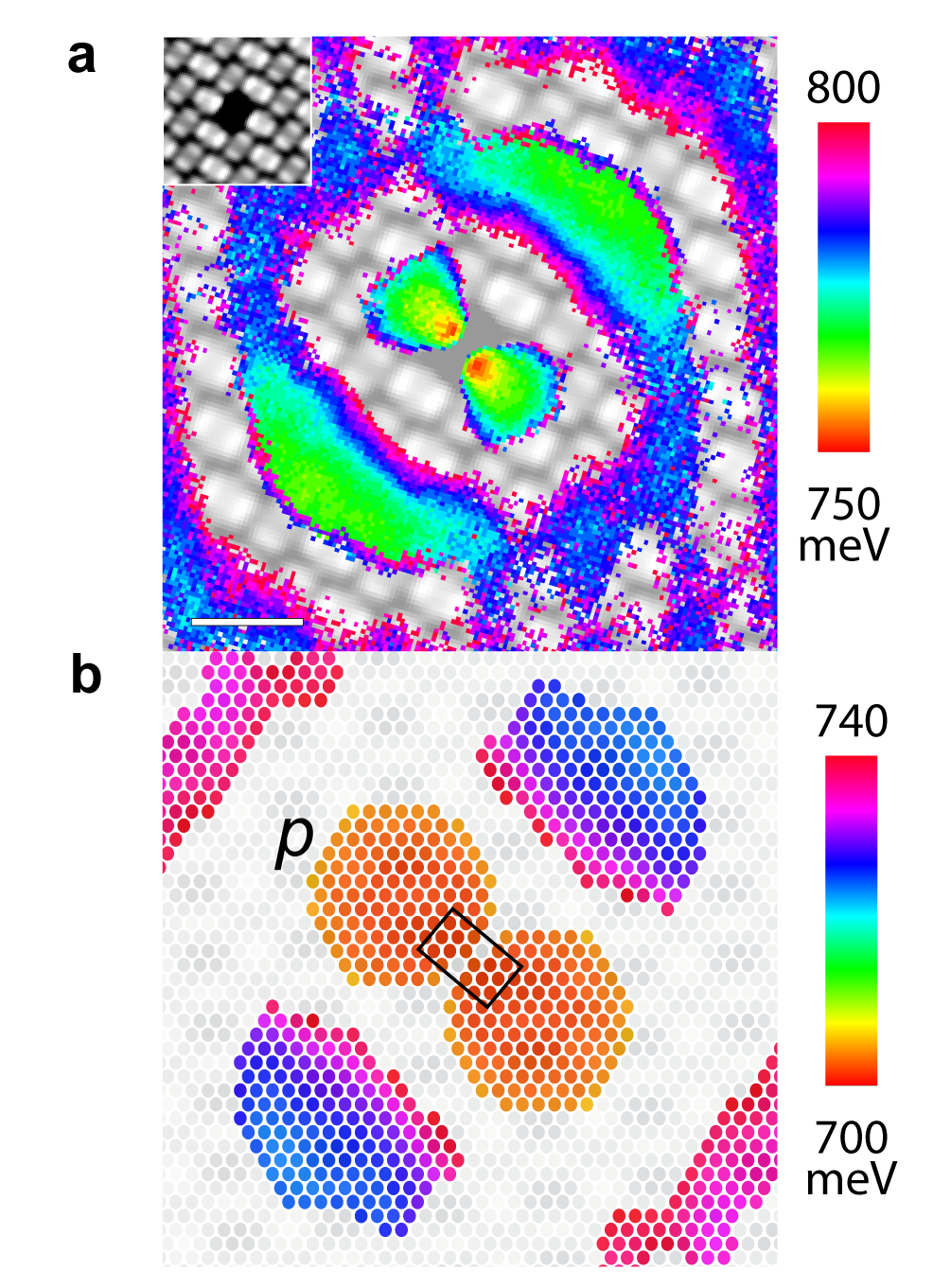}
\caption{Artificial p-orbital of a single molecular vacancy. 
(\textbf{a}) STM image of a type-B vacancy overlaid with a feature-detection map (FDM; see Methods) revealing the lowest-energy bound state of the vacancy; made from the portion of the peak-detection statistics marked with "p" in Fig.1e.
The colour reflects the peak energy, mapped according to the colorbar next to the image.
The inset provides a zoomed-out view of the vacancy, which in the main panel is covered by the FDM overlay.
(\textbf{b}) FDM obtained from the tight-binding (TB) simulation (see Methods).
The grey corrugation is due to the periodic potential (dark is more attractive).
The black rectangle was used to impose the vacancy potential.
The colour reflects the peak energy, mapped according to the colorbar next to the image.}
\label{vacB_FDSTS_p} 
\end{figure} 

\begin{figure}
\centering
\includegraphics[width=8cm]{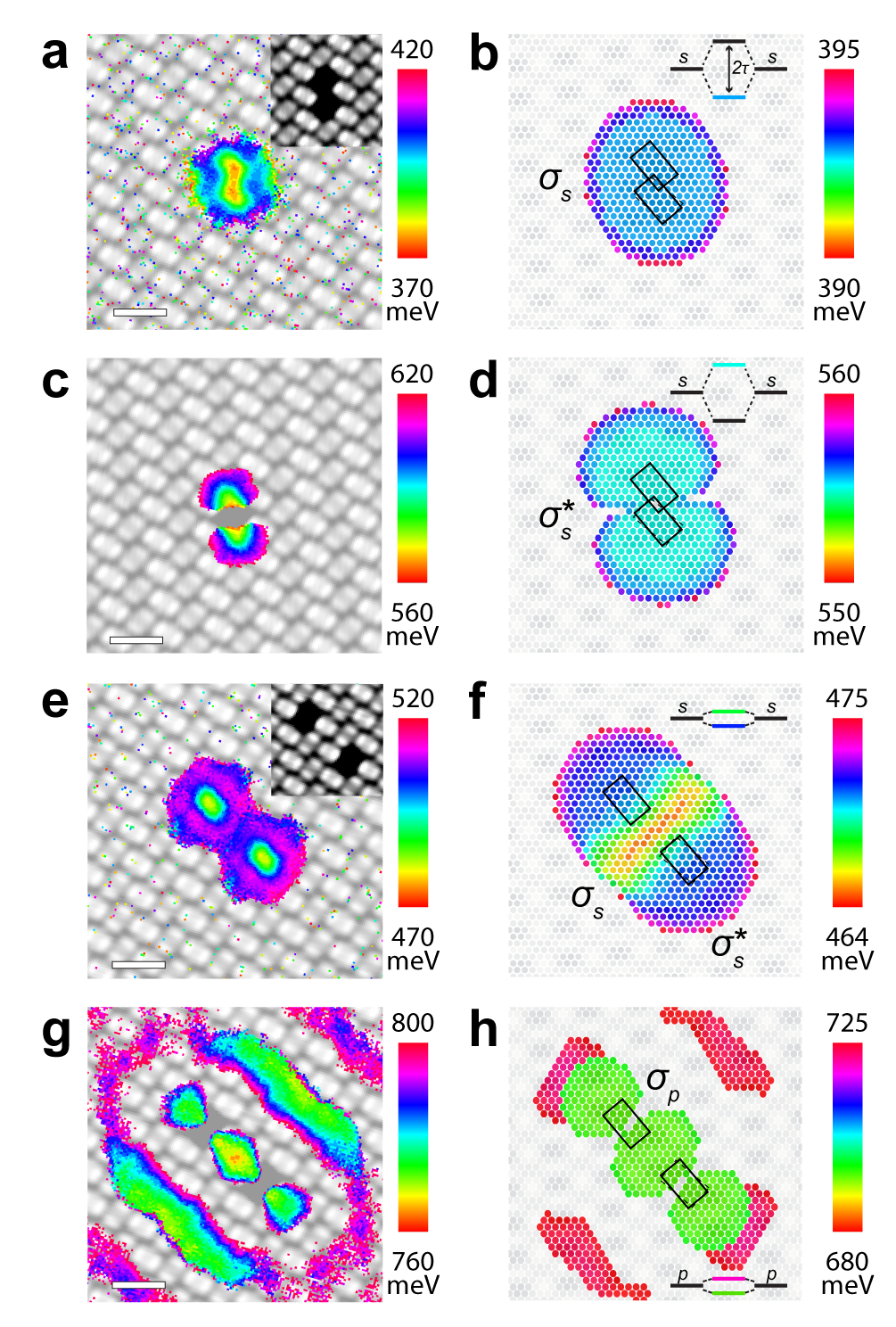}
\caption{Coupling of artificial orbitals in vacancy dimers. 
(\textbf{a--d}) Nearest-neighbour type-B vacancy pair: experimental (a,c) and simulated (b,d) feature-detection maps (FDMs; see Methods) showing the bonding (a,b) and antibonding (c,d) hybrids of two $s$ orbitals.
The schematic indicates the energy splitting $2\tau$ between the two hybrids.  
(\textbf{e--h}) A weekly interacting type-B vacancy pair: experimental (e,g) and TB simulated (f,h) FDMs highlighting the weak $s-s$ (e-f) and somewhat stronger $p-p$ coupling (g,h).
The black rectangles mark the model vacancy potential (see Methods).
The colour reflects the peak energy, mapped according to the colorbar next to the image.
The insets in (a,e) show the STM topographies of the dimers.
The insets in (b,d,f,h) indicate the degree of coupling and identify the shown hybrid orbital with its FDM colour.}

\label{vacBB1_FDSTS} 
\end{figure}

\begin{figure}
\centering
\includegraphics[width=8cm]{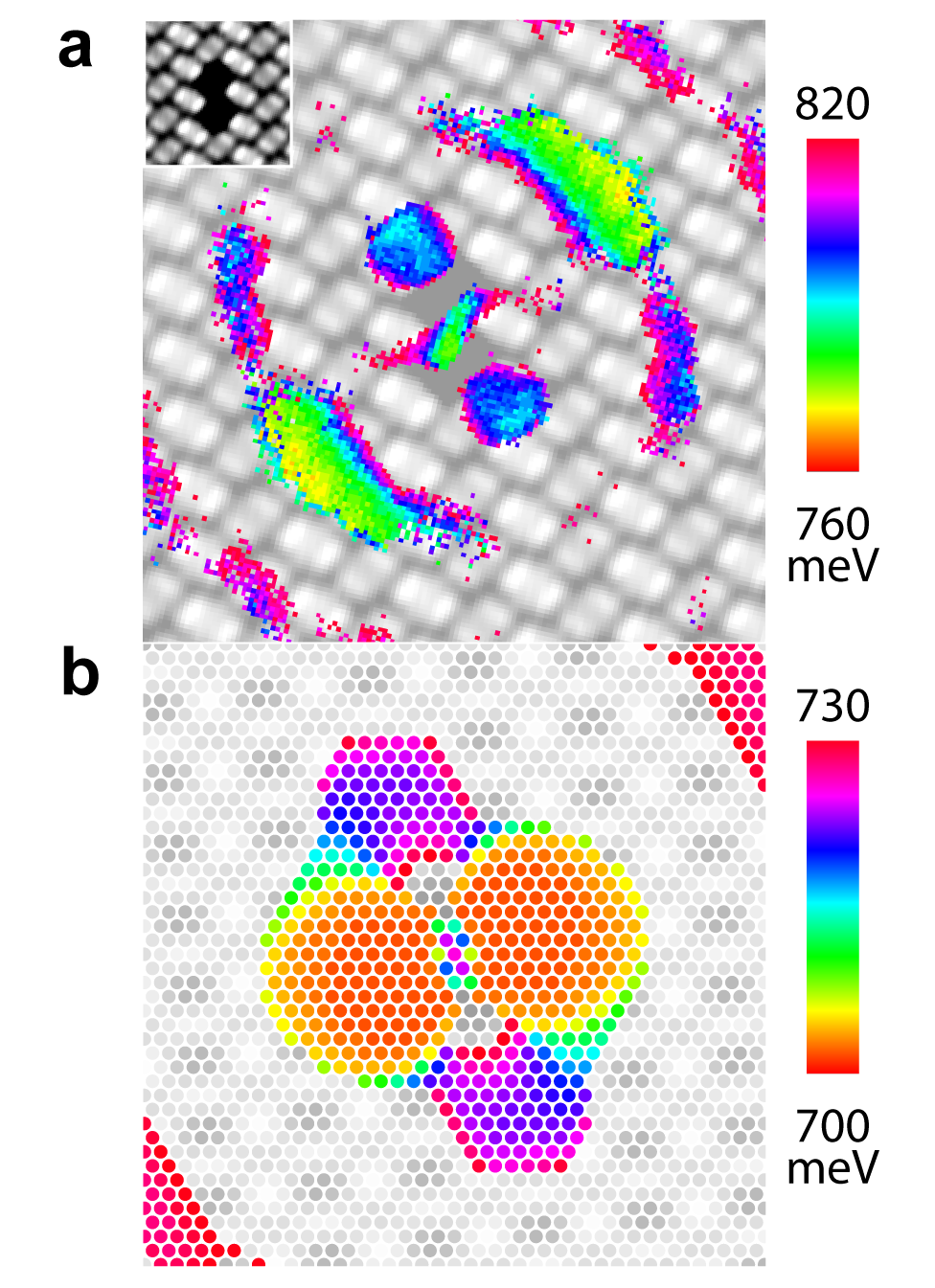}
\caption{Coupling of $p$ orbitals in the strongly interacting nearest-neighbour B vacancy pair. 
(\textbf{a}) STM image of a nearest-neighbour type-B vacancy pair overlaid with a feature-detection map (FDM; see Methods) showing a bound state that presumably arises due to the coupling of the $p$ states of individual vacancies (see Fig. 3 and Fig. 5 of the main text).
The colour scale indicates the energy of the detected spectral peak within the interval specified next to the image.
The inset provides a zoomed-out view of the vacancy area, otherwise partially covered by the FDM overlay.
(\textbf{b}) FDM made from the tight-binding (TB) data. 
Each coloured circle indicates the detection of a peak within the given energy interval. 
The gray corrugation represents the periodic potential of the PTCDA/Ag(111) interface, with darker regions corresponding to a more attractive potential (see Methods).
The colour scale indicates the energy of the detected spectral peak in the calculated FDM.}

\label{vacBB1_FDSTS_p} 
\end{figure}

\begin{figure}
\centering
\includegraphics[width=10cm]{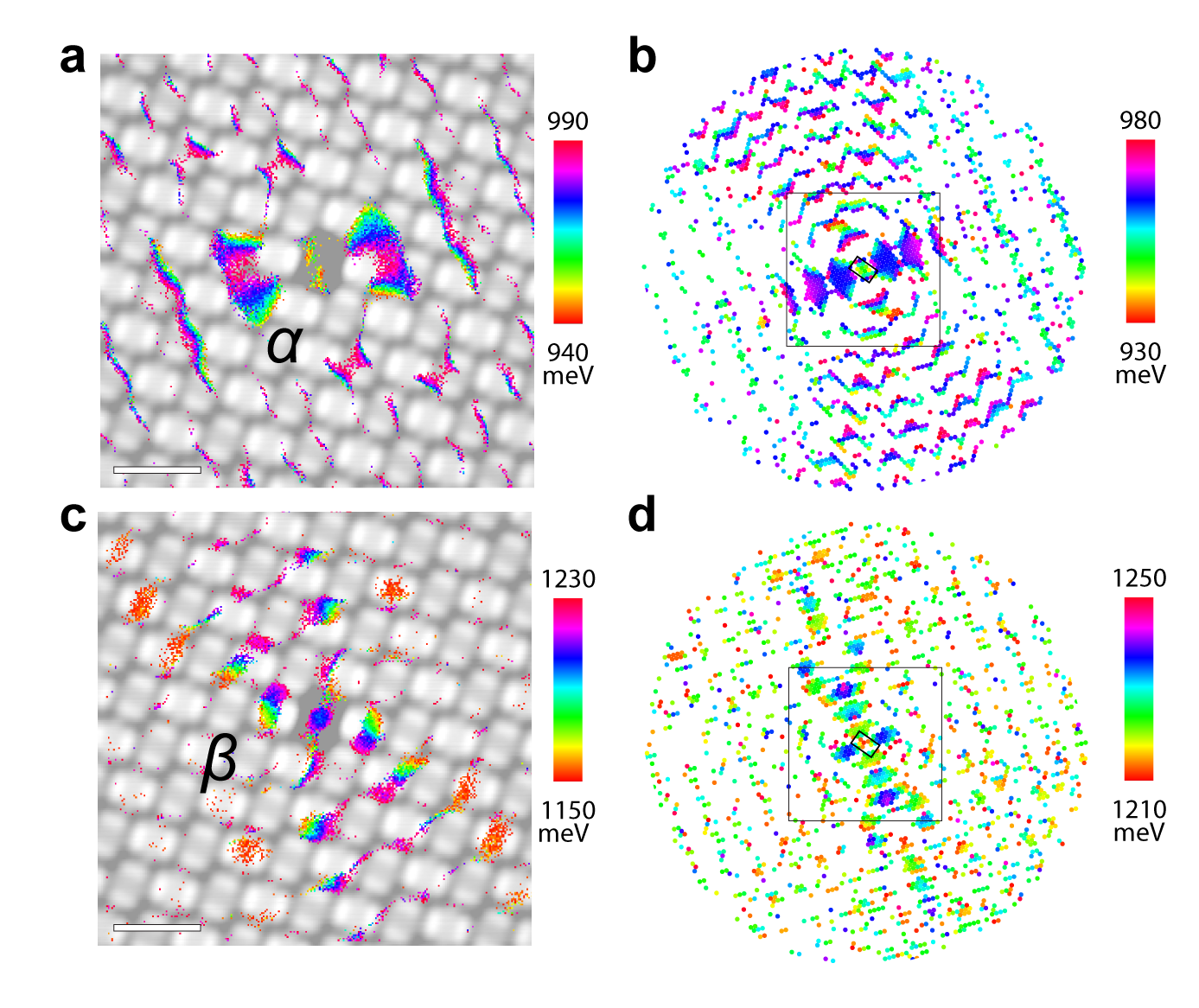}
\caption{High-energy quasi-localised orbitals of a single artificial atom. 
(\textbf{a,c}) STM image of a type-B vacancy overlaid with a feature-detection map (FDM; see Methods) revealing $\alpha$ and $\beta$ states;  made from the corresponding portions of the peak-detection statistics marked with "$\mathrm{\alpha}$" and "$\mathrm{\beta}$" in Fig.1e.
The colour reflects the peak energy, mapped according to the colorbar next to the image.
(\textbf{b,d}) Corresponding tight-binding (TB) simulated feature-detection maps (FDMs; see Methods).
The TB calculations were performed over an extended region to emphasise the localisation and wave-vector structure of these high-energy states.
The inner black rectangle marks the vacancy potential used in the model, and the outer square corresponds to the experimental field of view.
Colour scales indicate the energy of the detected spectral peaks within the specified intervals.}
\label{vacB_FDSTS_u} 
\end{figure}

\end {document}